\documentclass[pra,twocolumn,showpacs,preprintnumbers,amsmath,amssymb]{revtex4}

\usepackage{graphicx}
\usepackage{dcolumn}
\usepackage{bm}
\usepackage{color}

\begin{document}
\preprint{APS/123-QED}

\title{Exact Mobility Edges in One-Dimensional Mosaic Lattices Inlaid with Slowly Varying Potentials }
\author{Longyan Gong}
\thanks{Corresponding author. Email address:lygong@njupt.edu.cn}
\affiliation{
 $^{1}$College of Science, Nanjing University of Posts and Telecommunications, Nanjing, 210003, China\\
$^{2}$New Energy Technology Engineering of Jiangsu Province, Nanjing University of Posts and Telecommunications, Nanjing, 210003, China}

\date{today}
\begin{abstract}
We propose a family of one-dimensional mosaic models inlaid with a slowly varying potential $V_n=\lambda\cos(\pi\alpha n^\nu)$, where $n$ is the lattice site index and $0<\nu<1$. Combinating the asymptotic heuristic argument with the theory of trace map of transfer matrix, mobility edges (MEs) and pseudo-mobility edges (PMEs) in their energy spectra are solved semi-analytically, where ME separates extended states from weakly localized ones and PME separates weakly localized states from strongly localized ones. The nature of eigenstates in extended, critical, weakly localized and strongly localized is diagnosed by the local density of states, the Lyapunov exponent, and the localization tensor. Numerical calculation results are in excellent quantitative agreement with theoretical predictions.
\end{abstract}
\maketitle

\emph{Introduction.}---Anderson localization is one of the most focused phenomena in condensed matter physics~\cite{AN58,LA09,AB10}. In 1958, Anderson pointed out that disorder can destruct quantum interference and induce electron localization in three-dimensional random potential systems ~\cite{AN58}. As model parameters are varied, such as energy and disorder strength, a system can undergo phase transitions from the metallic phase with extended states to the insulator phase with localized states. When the disordered potential strength is below a threshold value, extended states are in the middle of band and localized states are near band edges. In the band, there are critical energies $E_c$, at which states being extended change to being localized. The critical energies are called mobility edges (MEs)~\cite{CO69,MO67}. It is known that at zero temperature, the conductivity would be vanish (finite) if the Fermi energy lies in a region of localized (extended) states. Therefore, the MEs can mark metal-insulator transitions or localization-delocalization transitions.

It is important to known what determines MEs. According to the scaling theory, all states are localized for one-dimensional (1D) Anderson model and there are no MEs~\cite{AB79}. At the same time, MEs are found in several 1D interesting models, e.g., the Soukoulis-Economou model with incommensurate potentials~\cite{SO82}, short-range correlated ~\cite{DU90} and long-range correlated ~\cite{DE98,IZ99} disordered potential model, and the Anderson model with long-range hopping\cite{RO03}. However, there are no analytical results about MEs for these models.

On the other hand, the Aubry-Andr\'{e}-Harper (AAH) model may be the most extensively studied 1D quasiperiodic model~\cite{AU80,HA55}. Using a duality transformation, extended (localized) states in position space can be mapped to localized (extended) ones in momentum space. There is a self-duality point that related to critical states. All states are extended, critical, or localized, which depend on potential strength. In other words, there exists a metal-insulator transition at a critical strength of the modulation potential. However, there are no MEs in its energy spectrum. By introducing the exponential short-range hopping terms, a modified AAH model has been proposed~\cite{BI10}. A family of generalized AAH models has also been investigated~\cite{GA15}. Based on generalized duality transformations, the analytical critical condition about MEs are predicted~\cite{BI10,GA15}. As Wang \emph{et al.} have pointed out, the self duality may be recovered on certain analytically determined energy, where the extended-localization transition occurs, while the whole model is not exactly solvable~\cite{WA20}. Based on the AAH model, they propose a class of 1D mosaic models and multiple MEs are analytically derived through computing the Lyapunov exponents (LEs) from Avila's global theory~\cite{WA20}. It is known that the LEs can well characterize the localization properties of exponential decaying states. However, for other types of wave function, e.g., power-law wave functions, the sole using of LEs is not sufficient and may lead to erroneous~\cite{VA92}. It will be interesting to exactly obtain MEs beyond the dual transformation and judging by the LEs. 

At the same time, the slowly varying potential model is another extensively studied 1D quasiperiodic model~\cite{GR88,TH88,DA88}. Its potential $V_n=\lambda\cos(\pi\alpha n^\nu)$, where $n$ is the lattice site index and $0<\nu<1$. Extended states have been found by the perturbation theory~\cite{GR88} and the WKB approximation~\cite{TH88}. An asymptotic semiclassical WKB-type theory predicts that two mobility edges are at $E_c=\pm(2-\lambda)$ for $\lambda\leq2$, extended states in the middle of the band ($|E|<2-\lambda$) and localized states at the band edge ($2-\lambda<|E|<2+\lambda$)~\cite{DA88,DA90}. At the same time, pseudo-mobility edges (PMEs) are found by a numerically accurate renormalization method~\cite{FA92,FA93}, which  separate weakly localized states from strongly localized ones.

It will be highly significant to develop generic models with multiple MEs and multiple PMEs. In this letter, inspired by the recent work~\cite{WA20}, we propose a family of 1D mosaic models inlaid with a slowly varying potential. They have richer phase diagrams than that for the slowly varying potential model and the mosaic ones based on AAH model. Using the asymptotic heuristic argument and the theory of trace map of transfer matrix, multiple MEs and multiple PMEs in their energy spectrum can be obtained semi-analytically. Further, our theory also predicts the localized and extended characterization of all states in energy spectrum.\\

\begin{figure}[!htbp]
\includegraphics[width=1.8in]{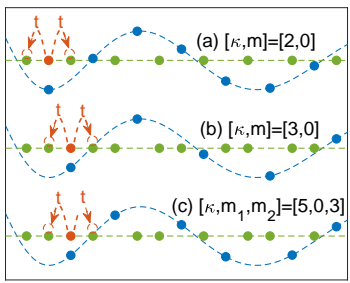}
\caption{The 1D quasiperiodic mosaic model with $[\kappa,m_1,m_2,...]=[2,0],[3,0]$ and $[5,0,3]$. The blue and green spheres denote the lattice sites whose potentials are quasiperiodic and zero, respectively, as shown by the corresponding blue and green dashed lines. The red sphere
denotes an electron, and the nearest-neighbor hopping strength is $t$.}\label{Fig1}
\end{figure}

\emph{Model.}---We consider an electron moving in 1D mosaic lattice models. The family of such models is described by
\begin{equation}
H=\sum\limits_{n}\lambda_n |n
\rangle\langle n|+t\sum\limits_{n} (|n \rangle\langle n+1|+|n+1 \rangle\langle n|)\label{EQ1}
\end{equation}
and
\begin{eqnarray}
\lambda_{n}=\left\{
\begin{array}
{r@{\quad\quad}l}$$\lambda\cos(\pi\alpha n^\nu),$~~~$ mod(n,\kappa)=m_1,m_2,...,\\
$$0,$~~~$\textrm{~~~~~~~~~~~~~~~~~~~~~~~~~otherwise},
\end{array}\right.\label{EQ2}
\end{eqnarray}
where $\lambda_n$ is the on-site potential on the $n$th site, $\lambda>0$ characterizes the quasiperiodic potential strength,
and $t$ is the nearest-neighbor hopping integral, which is used as energy unit. Further, $|n \rangle=c_n^{\dag}|0 \rangle$,
where $c_n^{\dag}$ is the creation operator of $n$th site and $|0 \rangle$ is the vacuum. Models are specified by the choice of $\kappa, m_1, m_2,...$, and we denote it by $[\kappa,m_1,m_2,...]$. It is the uniform potential model for $[\kappa,m]=[1,1]$ and it becomes the slowly varying potential model for $[\kappa,m]=[1,0]$.
The minimal nontrivial cases for $\kappa=2$ and $3$ are $[\kappa,m]=[2,0]$ and $[3,0]$, respectively. One of the minimal nontrivial cases for $\kappa=5$ with two $m$ is $[\kappa,m_1,m_2]=[5,0,3]$. Fig.\ref{Fig1} illustrates the model with $[\kappa,m_1,m_2,...]=[2,0],[3,0]$ and $[5,0,3]$, respectively, and other cases are similar.

\emph{Semi-analytical method.}---Now we provide the semi-analytical results about MEs and PEs by using the asymptotic heuristic argument~\cite{DA88,DA90} and the theory of trace map of transfer matrix~\cite{KO83}. The slowly varying potential $V_n=\lambda\cos(\pi\alpha n^\nu)$ given in Eq.(\ref{EQ2}) has highly correlated disorder feature~\cite{IZ99}. It can be written as
\begin{equation}
\frac{dV_n}{dn}=-\lambda\pi\alpha{n}^{\nu-1}\sin(\pi\alpha{n}^{\nu}).\label{EQ3}
\end{equation}
When $n\to\infty$ and $0<\nu<1$,
\begin{equation}
\bigg|\frac{dV_n}{dn}\bigg|\sim \frac{|\sin(\pi\alpha{n}^{\nu})|}{n^{1-\nu}}\rightarrow0.\label{EQ4}
\end{equation}
Equivalently, $V_{n+1}-V_{n}\sim O(|n|^{\nu-1})$, which vanishes for large $n$. Therefore, such potential can be taken as a ``local constancy''~\cite{DA88,DA90}.

On the other hand, the Schrd\"{o}inger equation for the Hamiltonian given in Eq.(\ref{EQ1}) can be written as
\begin{equation}
t\phi_{n-1}+t\phi_{n+1}+\lambda_n\phi_{n}=E\phi_{n},\label{EQ5}
\end{equation}
where $\phi_{n}$ is the amplitude of wave at $n$th site and $E$ is the corresponding eigenenergy. Eq.(\ref{EQ5}) can be rewritten in terms of the transfer matrix $M(n)$,
\begin{equation}
\Phi_{n+1}=\left(
  \begin{array}{ccc}
    \frac{E-\lambda_n}{t}& -1\\
     1                   &  0\\
  \end{array}
\right)
\Phi_{n}
\equiv M(n) \Phi_{n}, \label{EQ6}
\end{equation}
where
\begin{equation}
\Phi_n=\left(
  \begin{array}{ccc}
    \phi_n\\
    \phi_{n-1}\\
  \end{array}
\right). \label{EQ7}
\end{equation}
From Eq.(\ref{EQ2}), $V_n=\lambda\cos(\pi\alpha n^\nu)$ periodically occurs with interval $\kappa$, so we can introduce a quasicell with the nearest $\kappa$ lattice sites.
In a quasicell, by a successive application Eq.(\ref{EQ6}), a recursion relation is given by
\begin{equation}
\Phi_{n+1+\kappa}=M(n+\kappa)...M(n+1) \Phi_{n+1}\equiv M^\kappa(n)\Phi_{n+1}. \label{EQ8}
\end{equation}
If there are $j$ same quasicells, the relation is
\begin{equation}
\Phi_{n+1+j\kappa}=[M^\kappa(n)]^j\Phi_{n+1}. \label{EQ9}
\end{equation}
Let
\begin{equation}
\chi(E)=tr M^\kappa(n)/2, \label{EQ10}
\end{equation}
then $|\chi|<1$ for allowed energies~\cite{KO83}.

For the slowly varying potential $V_n$, $V_{max}=\lambda$ and $V_{min}=-\lambda$, which correspond to the maximum potential barrier and potential well.
As $n\to\infty$, $V_n$ is a ``local constancy''~\cite{DA88,DA90}. Near the maximum potential barrier and potential well, we respectively use Eq.(\ref{EQ10}), and corresponding $V_n$ are approximated by $\lambda$ and $-\lambda$. For the two cases, we denote $\chi(E)$ in Eq.(\ref{EQ10}) by $\chi_\lambda(E)$ and $\chi_{-\lambda}(E)$, respectively. Our rule to judge the localization properties of states is summarized in Table I, which is the main result in this paper. Table I means that an electron with $E$ is in an extended state if it can penetrate both the maximum potential barrier and the maximum potential well. It is in a weakly localized state if it can penetrate only one. It is in a strongly localized state if it penetrates neither. It is in a critical state if it just penetrates one or both, therefore MEs and PMEs are determined by the functions $\chi_{\lambda}(E)=\pm1$ or $\chi_{-\lambda}(E)=\pm1$.

\begin{table}[tp]
	\caption{\label{tab}%
		State properties judging by $\chi_{\lambda}(E)$ and $\chi_{-\lambda}(E)$, where $E_x$, $W$, $S$ and $C$ represent extended, weakly localized, strongly localized and critical, respectively.
	}
	\begin{ruledtabular}
		\begin{tabular}{ccc}
			$|\chi_{\lambda}(E)|$ & $|\chi_{-\lambda}(E)|$ &\textrm{state property}\\
			\colrule
			$<1$ & $<1$  & $E_x$\\
			$<1$ & $>1$  & $W$  \\
			$>1$ & $<1$  & $W$  \\
            $>1$ & $>1$  & $S$  \\
			$=1$ & $>1,=1,<1$  & $C$  \\
			$>1,=1,<1$ & $=1$  & $C$  \\
		\end{tabular}
	\end{ruledtabular}
\end{table}

\emph{Phase diagram.}---For model parameters $[\kappa,m_1,m_2,...]=[1,1],[1,0],[2,0],[3,0]$ and $[5,0,3]$, base on Eq.(\ref{EQ10}), we get
\begin{equation}
\chi_{\lambda}^{[1,1]}=\chi_{-\lambda}^{[1,1]}=\frac{E}{2}, \label{EQ11}
\end{equation}

\begin{eqnarray}
\left\{
\begin{array}
{r@{\quad\quad}l}$$\chi_{\lambda}^{[1,0]}$=$\frac{1}{2}(E-\lambda),\\
$$\chi_{-\lambda}^{[1,0]}                $=$\frac{1}{2}(E+\lambda),
\end{array}\right.\label{EQ12}
\end{eqnarray}

\begin{eqnarray}
\left\{
\begin{array}
{r@{\quad\quad}l}$$\chi_{\lambda}^{[2,0]}$=$\frac{1}{2}(E^2-\lambda E-2),\\
$$\chi_{-\lambda}^{[2,0]}                $=$\frac{1}{2}(E^2+\lambda E-2),
\end{array}\right.\label{EQ13}
\end{eqnarray}

\begin{eqnarray}
\left\{
\begin{array}
{r@{\quad\quad}l}$$\chi_{\lambda}^{[3,0]}$=$\frac{1}{2}(E^3-\lambda E^2-3E+\lambda),\\
$$\chi_{-\lambda}^{[3,0]}               $=$\frac{1}{2}(E^3+\lambda E^2-3E-\lambda),
\end{array}\right.\label{EQ14}
\end{eqnarray}
and
\begin{eqnarray}
\left\{
\begin{array}
{r@{\quad\quad}l}$$\chi_{\lambda}^{[5,0,3]}$=$\frac{1}{2}[E^5-2\lambda E^4+(\lambda^2-5)E^3\\+6\lambda E^2-(\lambda^2-5)-2\lambda],\\
$$\chi_{-\lambda}^{[5,0,3]}                $=$\frac{1}{2}[E^5+2\lambda E^4+(\lambda^2-5)E^3\\-6\lambda E^2-(\lambda^2-5)+2\lambda].
\end{array}\right.\label{EQ15}
\end{eqnarray}

Judged by the rule in Table I, for $[\kappa,m]=[1,1]$, states with $|E|<2$ are extended, which are just state localization properties for the uniform potential model. For $[\kappa,m]=[1,0]$, extended states in the middle of the band ($|E|<2-\lambda$) at $\lambda\leq2$, localized states at the band edge ($2-\lambda<|E|<2+\lambda$), and
two MEs are at $E_c=\pm(2-\lambda)$, which are just that obtained for the slowly varying potential model in Refs.~\cite{DA88,DA90}.

\begin{figure}[!htbp]
\includegraphics[width=1.85in,height=1.2in]{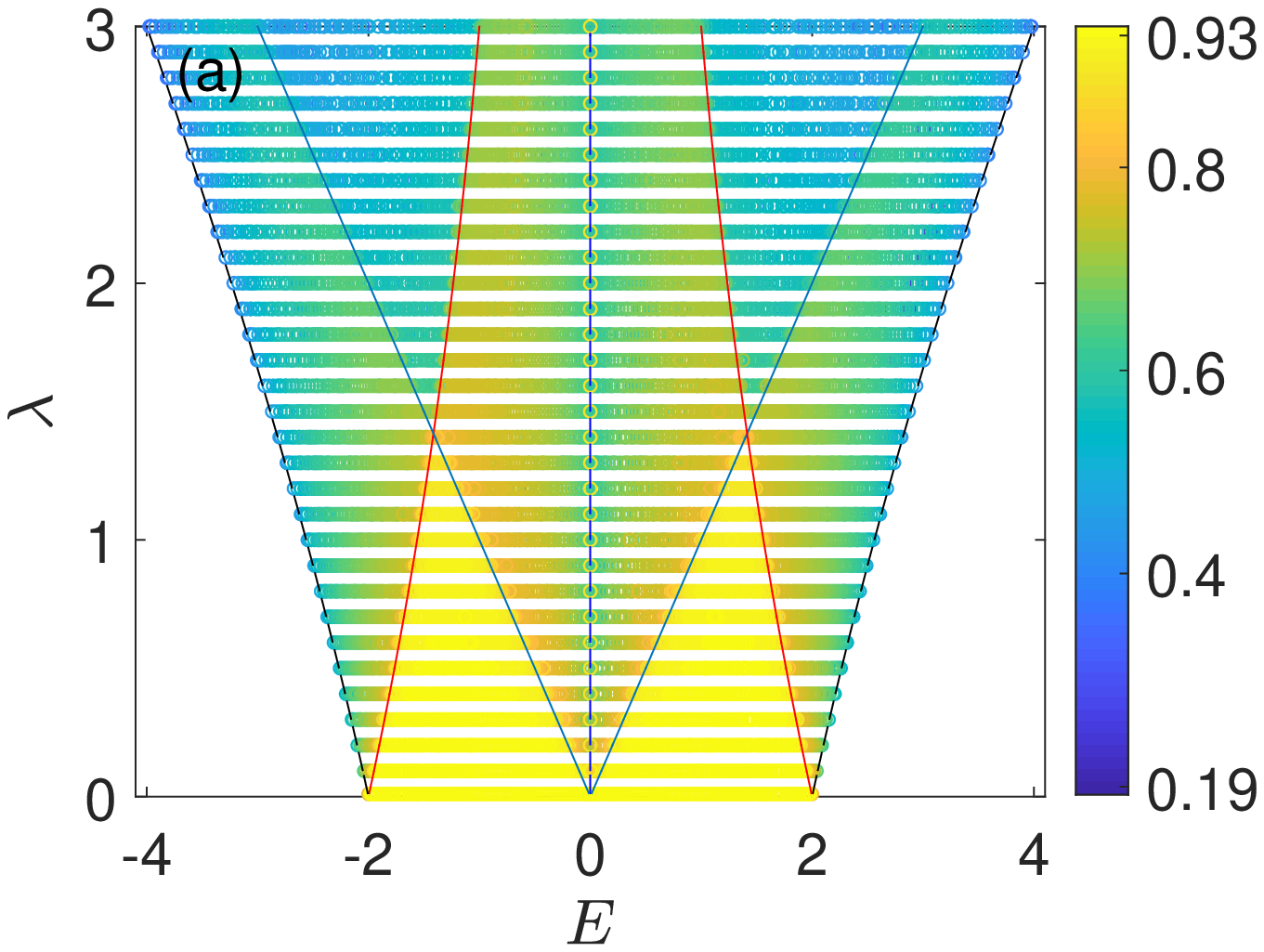}~~
\includegraphics[width=1.5in]{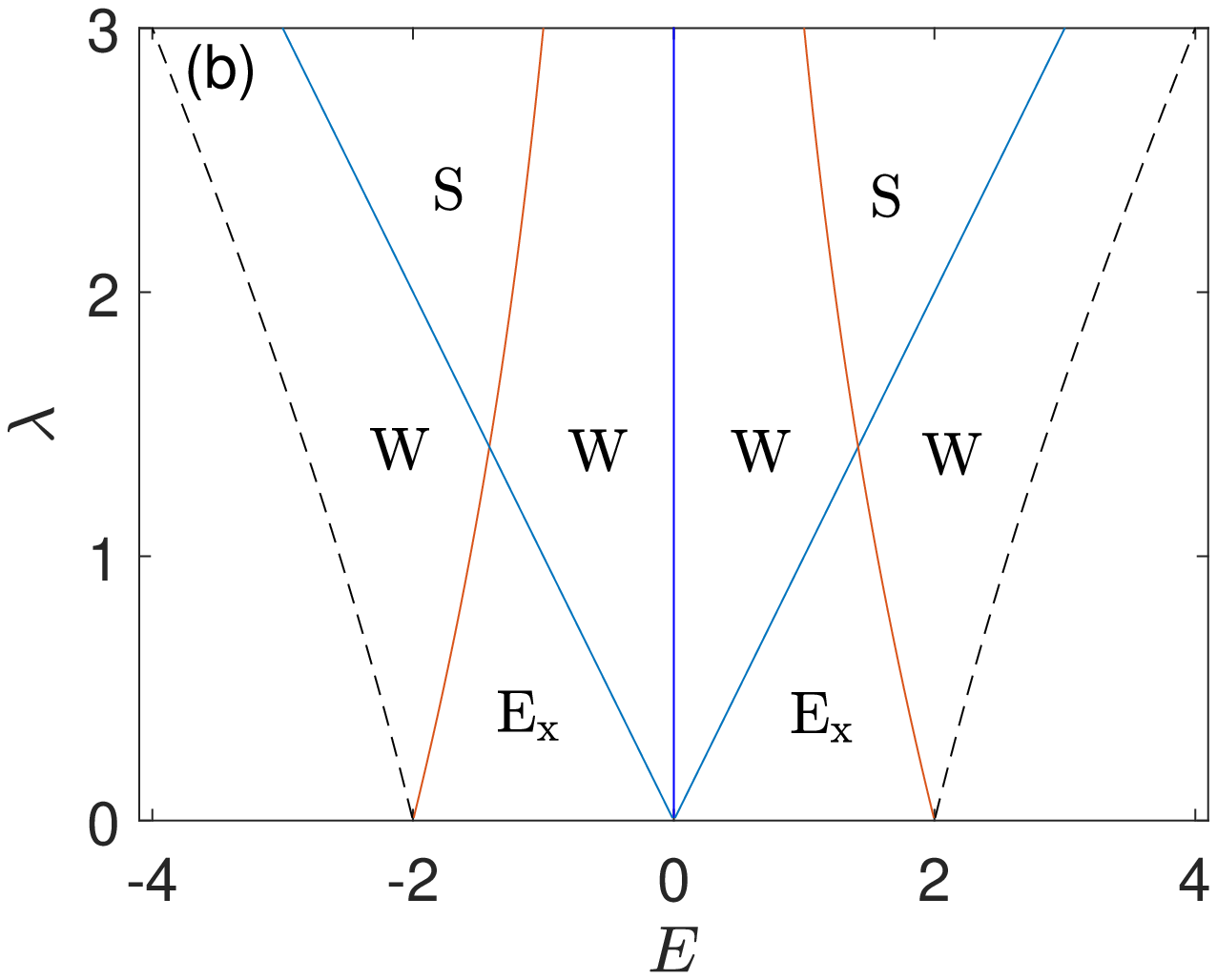}
\caption{For $[\kappa,m]=[2,0]$, (a)fractal dimension $\Gamma$ as functions of eigenenergies $E$ and
potential strength $\lambda$, where the system size $N=500$, $\pi\alpha=0.2$ and $\nu=0.7$, respectively. (b) Phase diagram. The lines in (a) and (b) represent MEs and PMEs, and $E_x$, $W$, and $S$ respectively represent extended, weakly localized, and strongly localized phases.}\label{Fig2}
\end{figure}

\begin{figure}[!htbp]
\includegraphics[width=1.85in,height=1.2in]{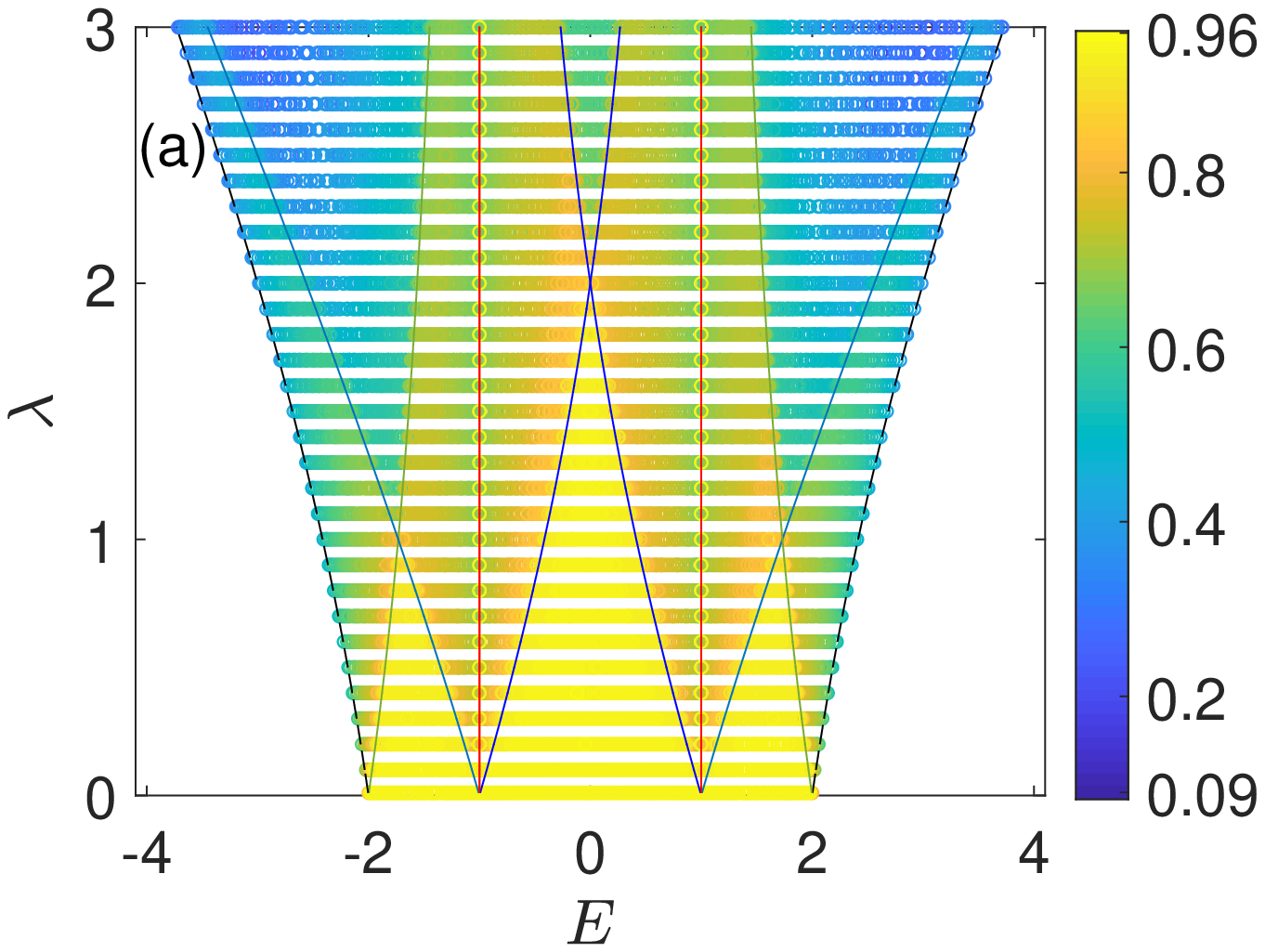}~~
\includegraphics[width=1.5in]{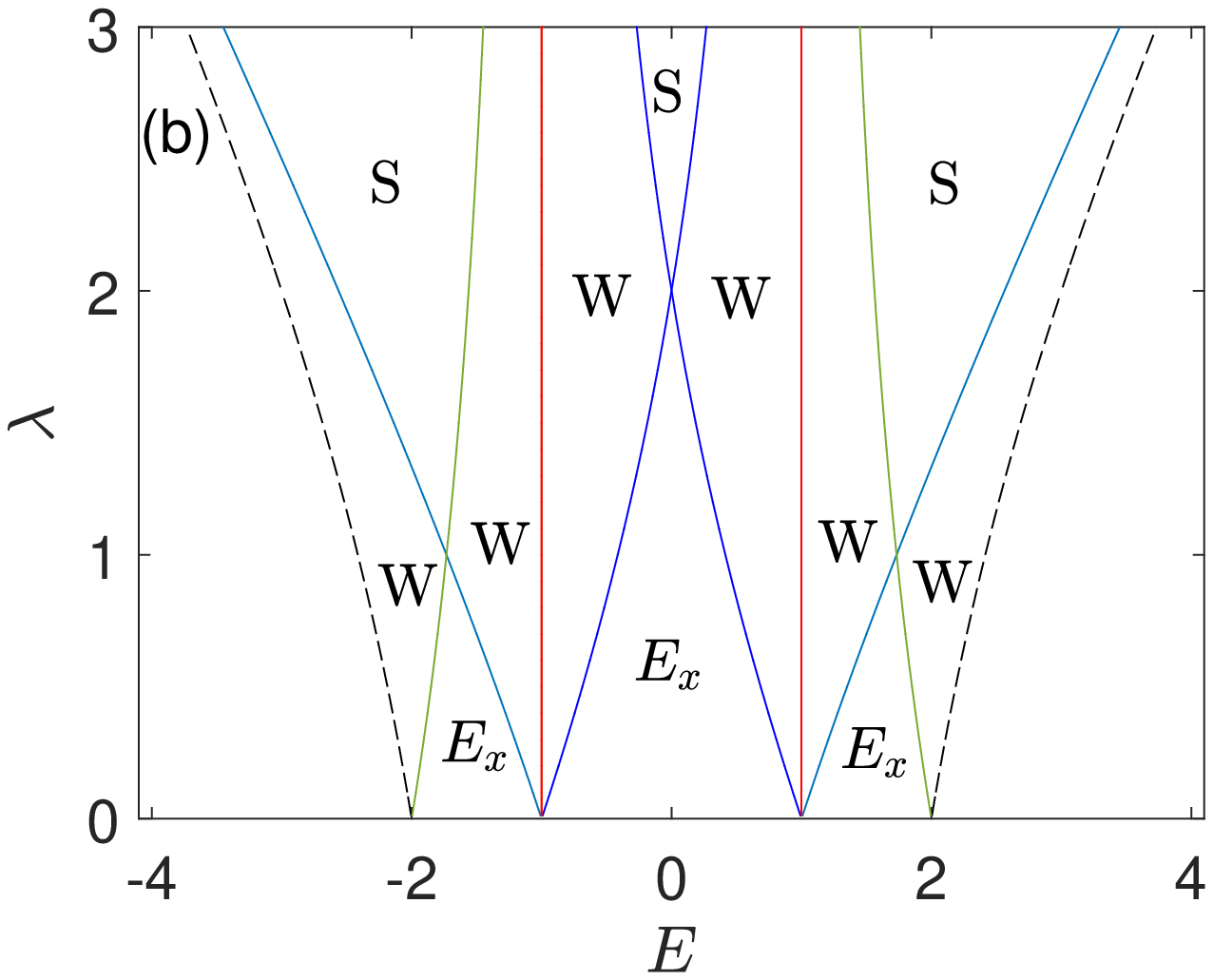}
\caption{The same as Fig.\ref{Fig2}, but $[\kappa,m]=[3,0]$ and system size $N=600$.}\label{Fig3}
\end{figure}

\begin{figure}[!htbp]
\includegraphics[width=1.85in,height=1.2in]{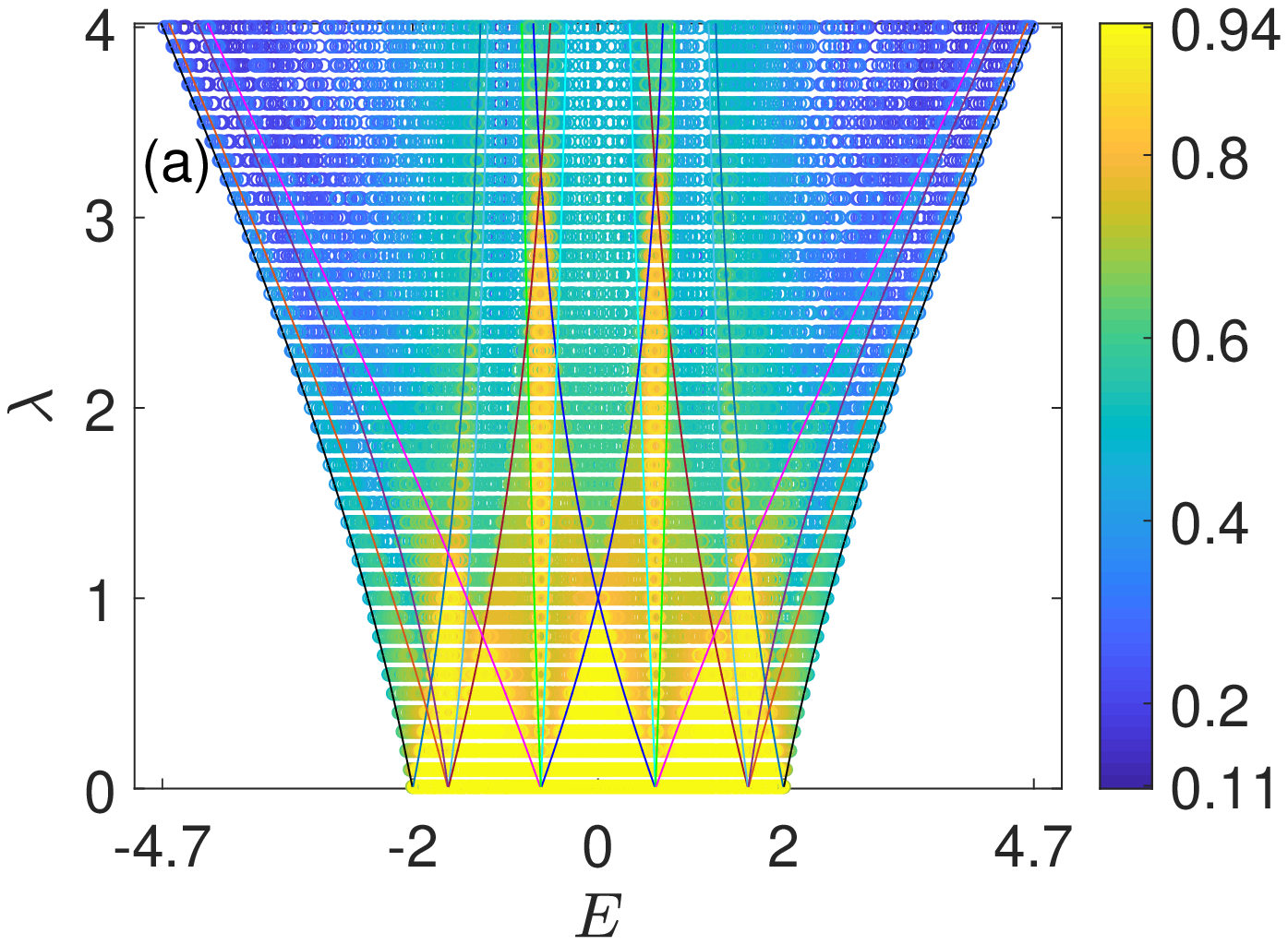}~~
\includegraphics[width=1.5in]{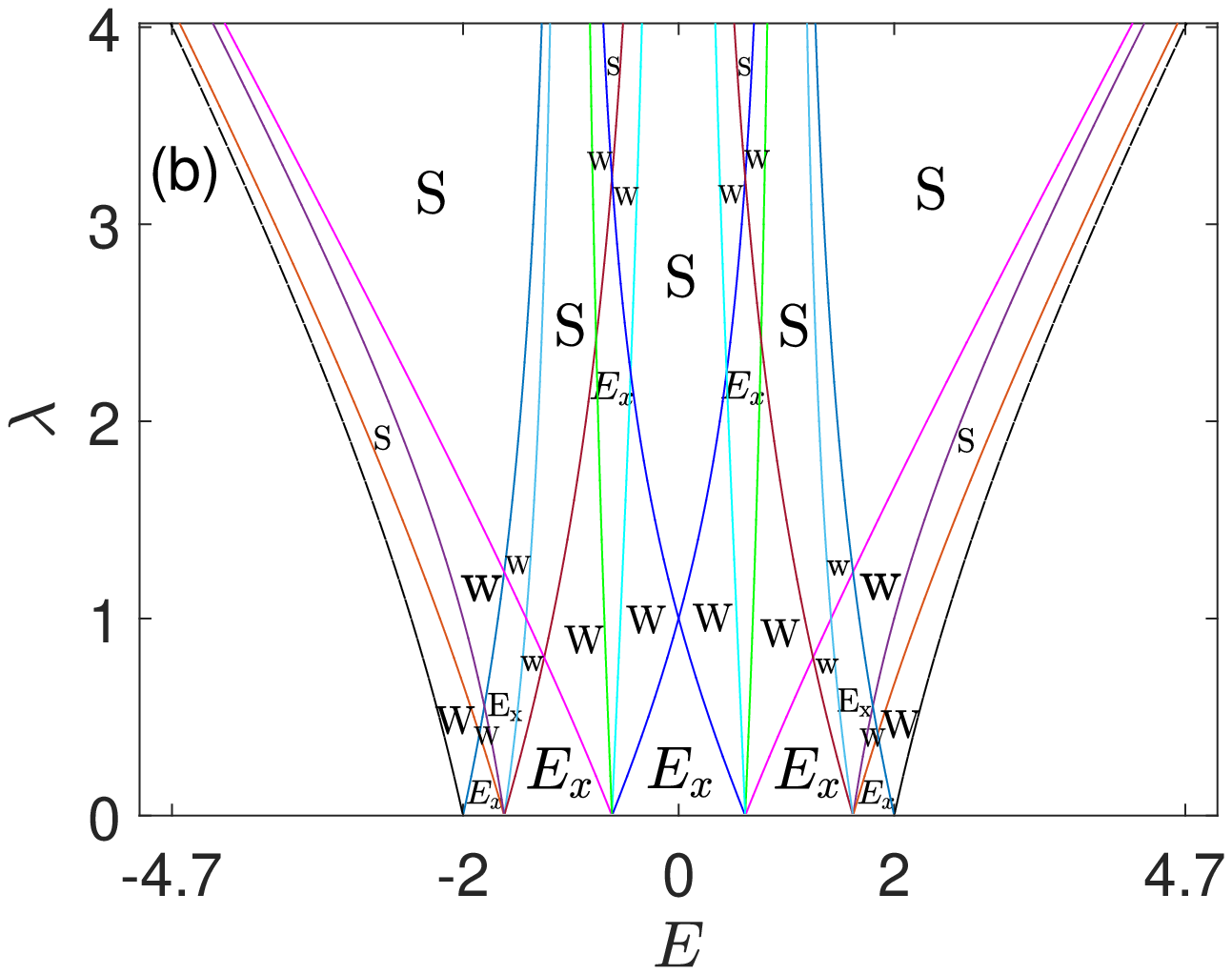}
\caption{The same as Fig.\ref{Fig2}, but $[\kappa,m_1,m_2]=[5,0,3]$.}\label{Fig4}
\end{figure}

To illustrate phase diagrams for $[\kappa,m_1,m_2,...]=[2,0],[3,0]$ and $[5,0,3]$, we study the fractal dimension $\Gamma$ of eigenfunction $\phi_{\beta}$, where $\beta$ is the index of eigenfunction. The fractal dimension
\begin{equation}
\Gamma=-\lim_{N\to\infty}[\ln(IPR)/\ln N], \label{EQ16}
\end{equation}
where the inverse participation ratio $IPR=\sum_{n=1}^N|\phi_{\beta,n}|^4$\cite{EV08}.  For $[\kappa,m_1,m_2,...]=[2,0],[3,0]$ and $[5,0,3]$, $\Gamma$ as functions of eigenenergies $E$ and potential strength $\lambda$ are plotted in Figs.\ref{Fig2}(a), \ref{Fig3}(a) and \ref{Fig4}(a), respectively. The lines in them represent the MEs and PMEs, which are determined with Eqs.(\ref{EQ13})-(\ref{EQ15}) by the conditions that $\chi_{\lambda}(E)=\pm1$ or $\chi_{-\lambda}(E)=\pm1$. According to the rule in Table I, extended ($E_x$), weakly localized ($W$) and stronger localized ($S$) phases are labelled in Fig.\ref{Fig2}(b), \ref{Fig3}(b) and \ref{Fig4}(b), respectively.
There are MEs separating extended states from weakly localized states. There also exist PMEs separating weakly localized states from strongly localized states. We numerically find the two outermost MEs are beyond the allowed energies, which are denoted by dashed lines. It is known that $\Gamma\to 1$ for extended states and $\Gamma\to 0$ for localized ones. As expected from analytical results, $\Gamma$ is less than and close to one for states in extended phases, approaches zero for states in stronger localized phases, and it has intermediate values for states in weakly localized phases.

From Eq.(\ref{EQ10}) and these obviously displayed in Eqs.(\ref{EQ11})-(\ref{EQ15}), we know the trace $\chi(E)$ is a $\kappa$th-order polynomial of $E$. At the same time, MEs and PMEs are determined by the conditions that $\chi_{\lambda}(E)=\pm1$ and $\chi_{-\lambda}(E)=\pm1$. Therefore, the number of MEs and PMEs $N_{m}\leq4\kappa$, and $N_{m}=4\kappa$ if all roots are different, so there are $N_m-1$ different phases. 
\begin{figure}[!htbp]
\includegraphics[width=1.63in]{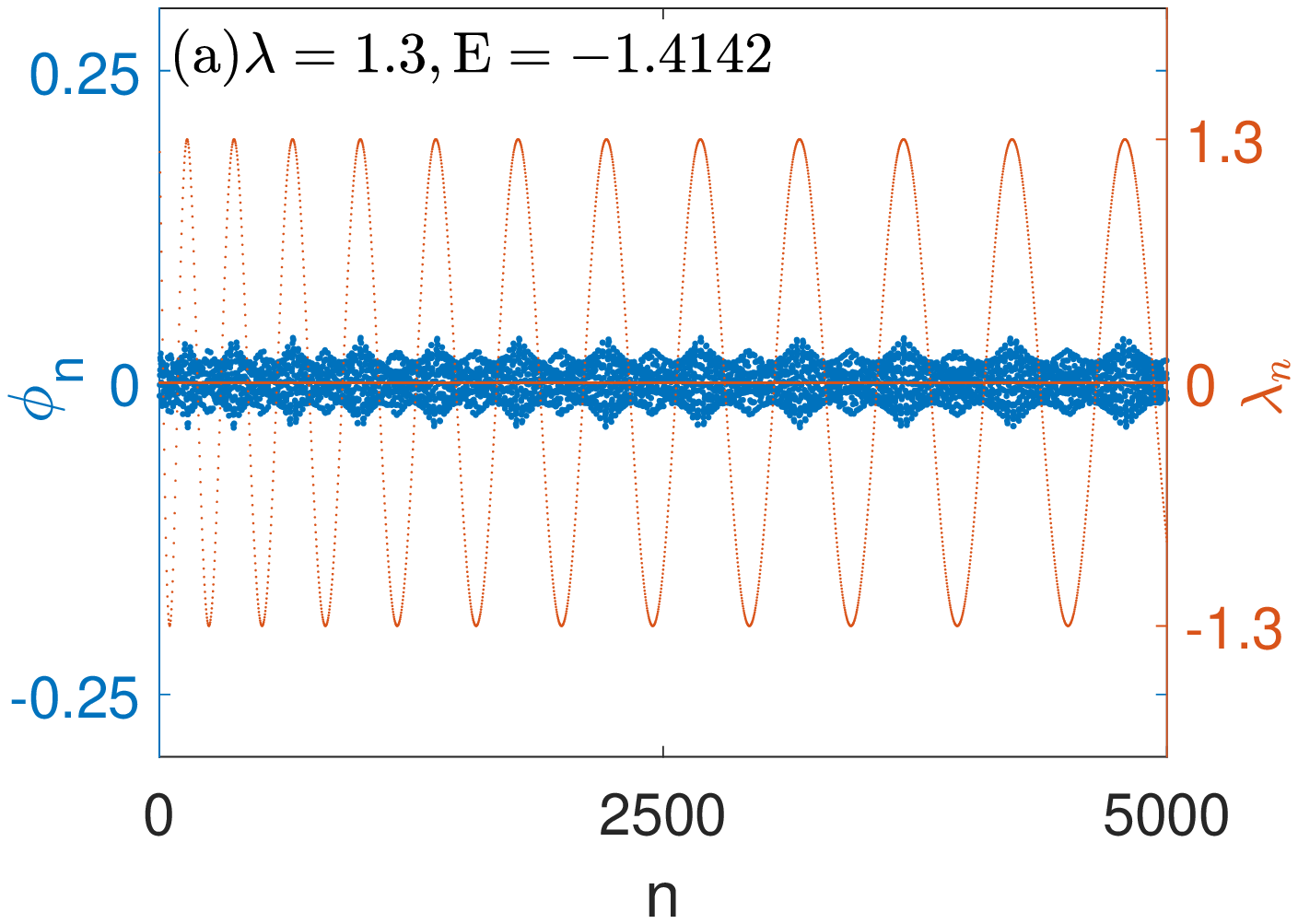}~
\includegraphics[width=1.69in]{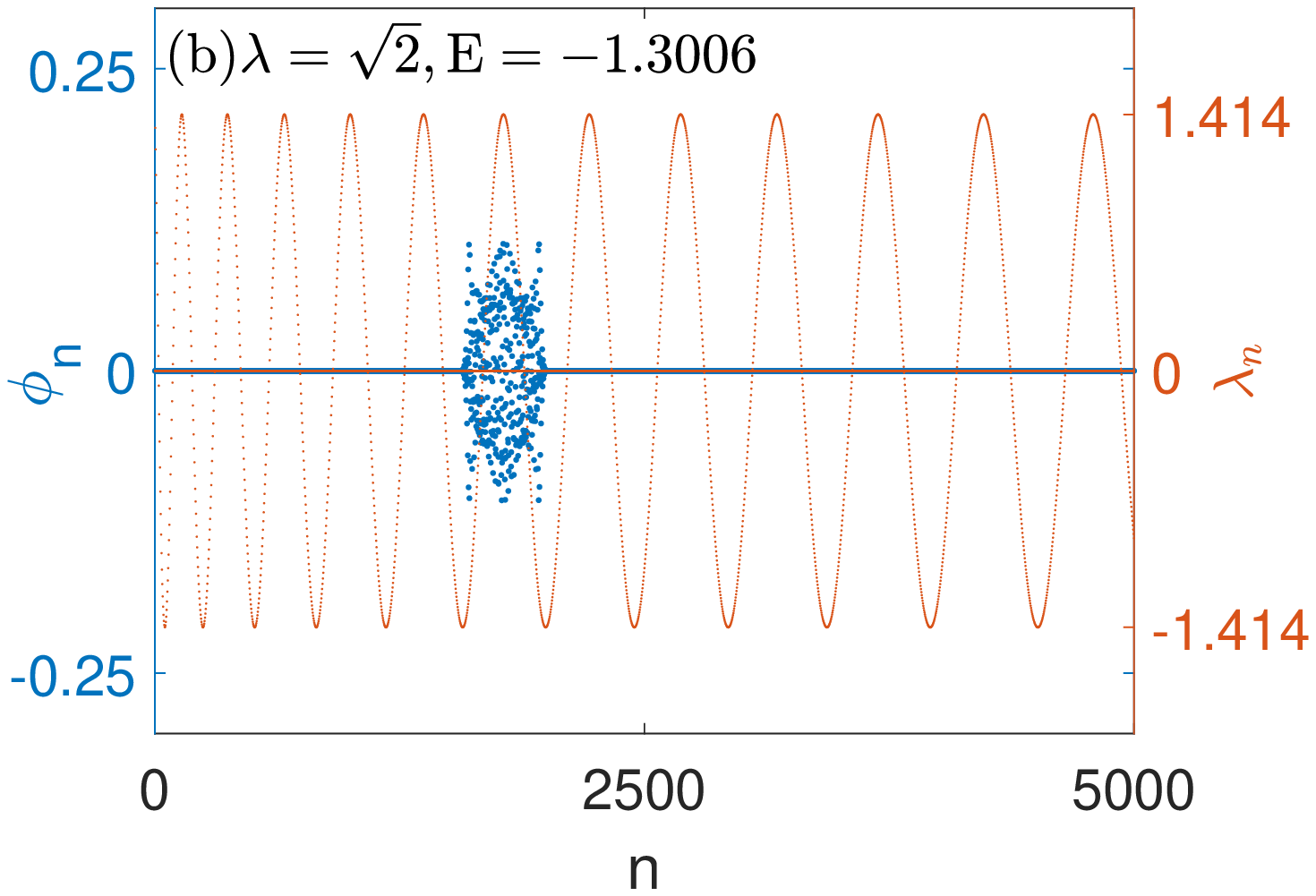}
\vspace{0.3cm}

\includegraphics[width=1.63in]{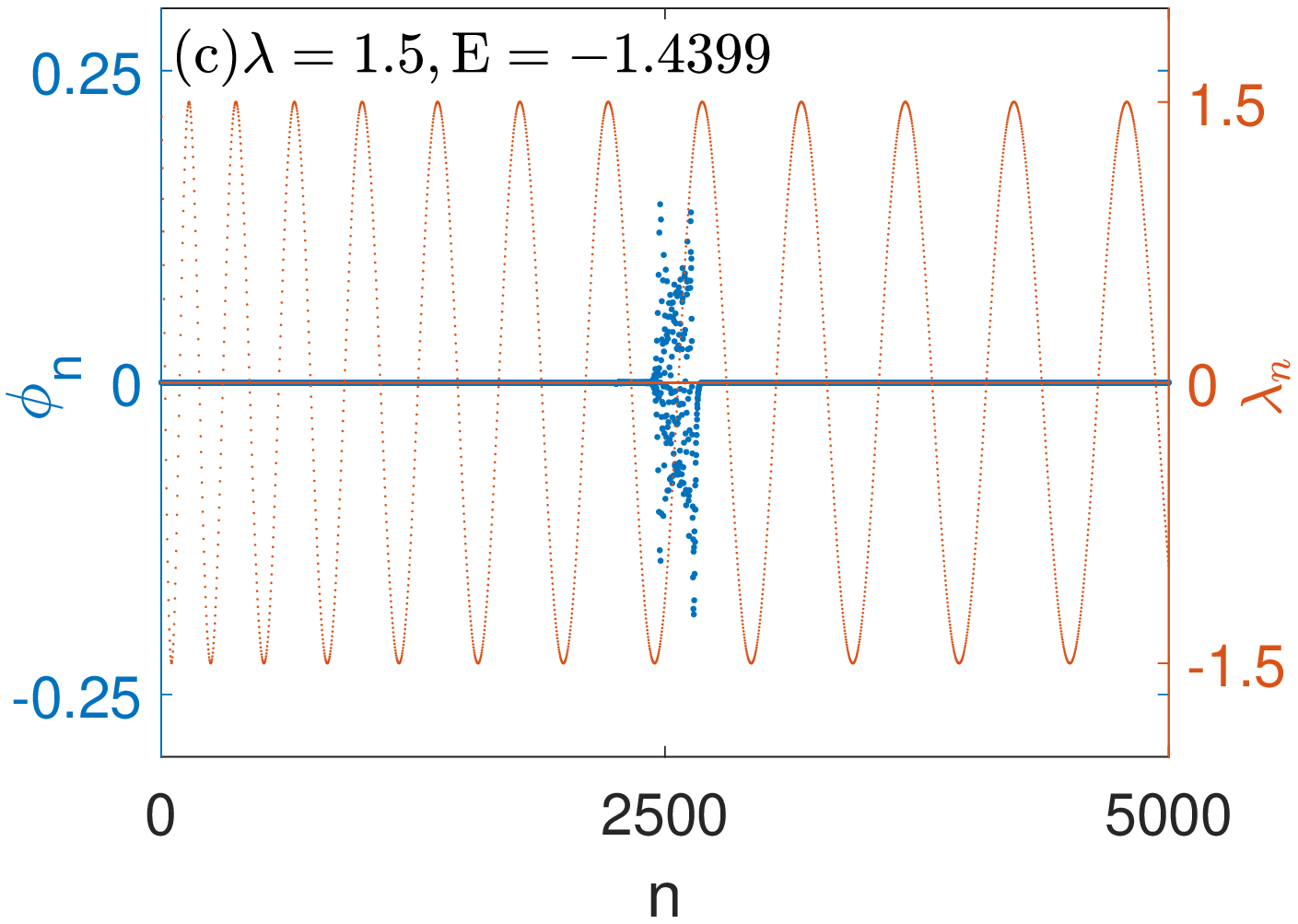}~
\includegraphics[width=1.63in]{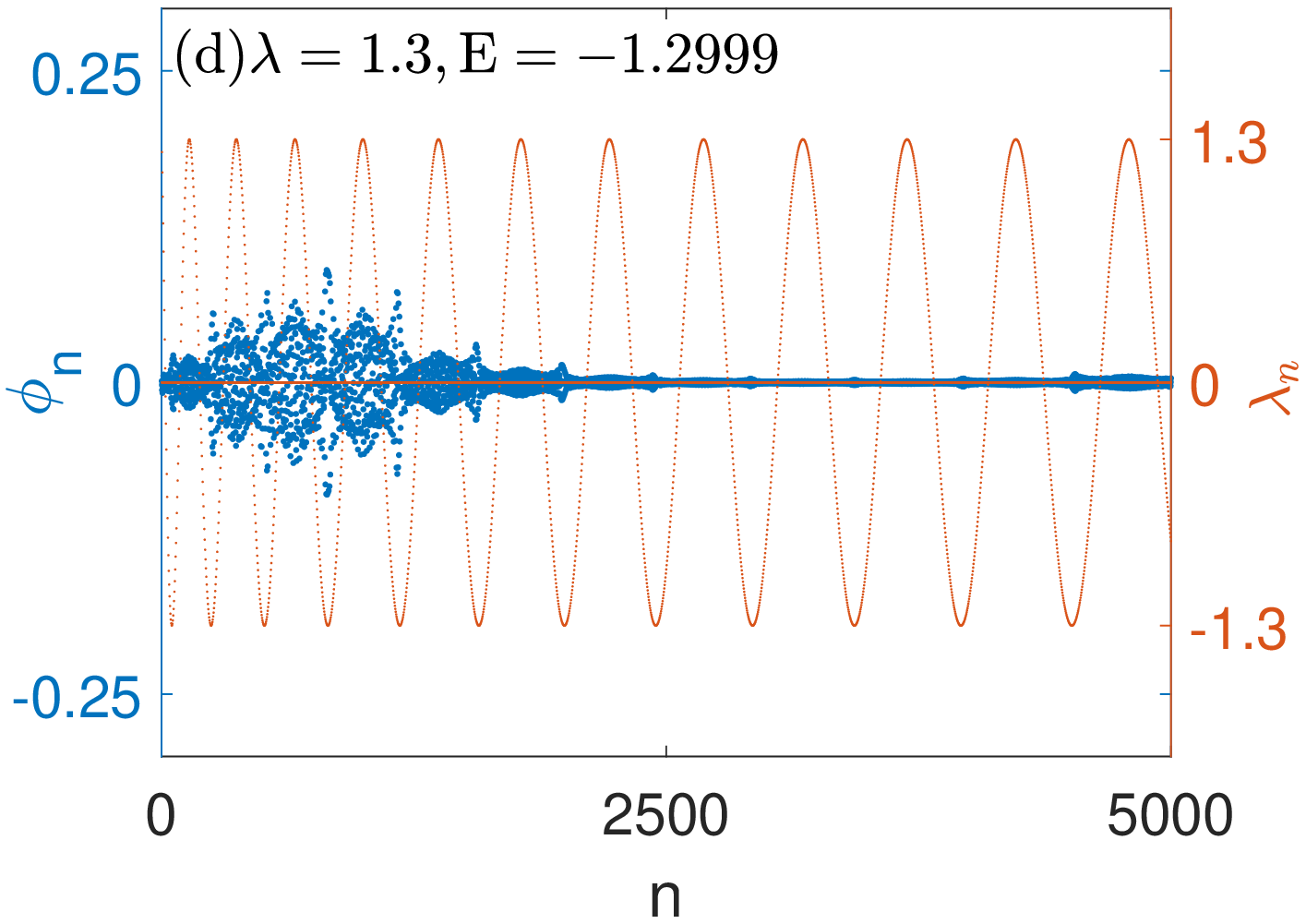}
\vspace{0.3cm}

\includegraphics[width=1.63in]{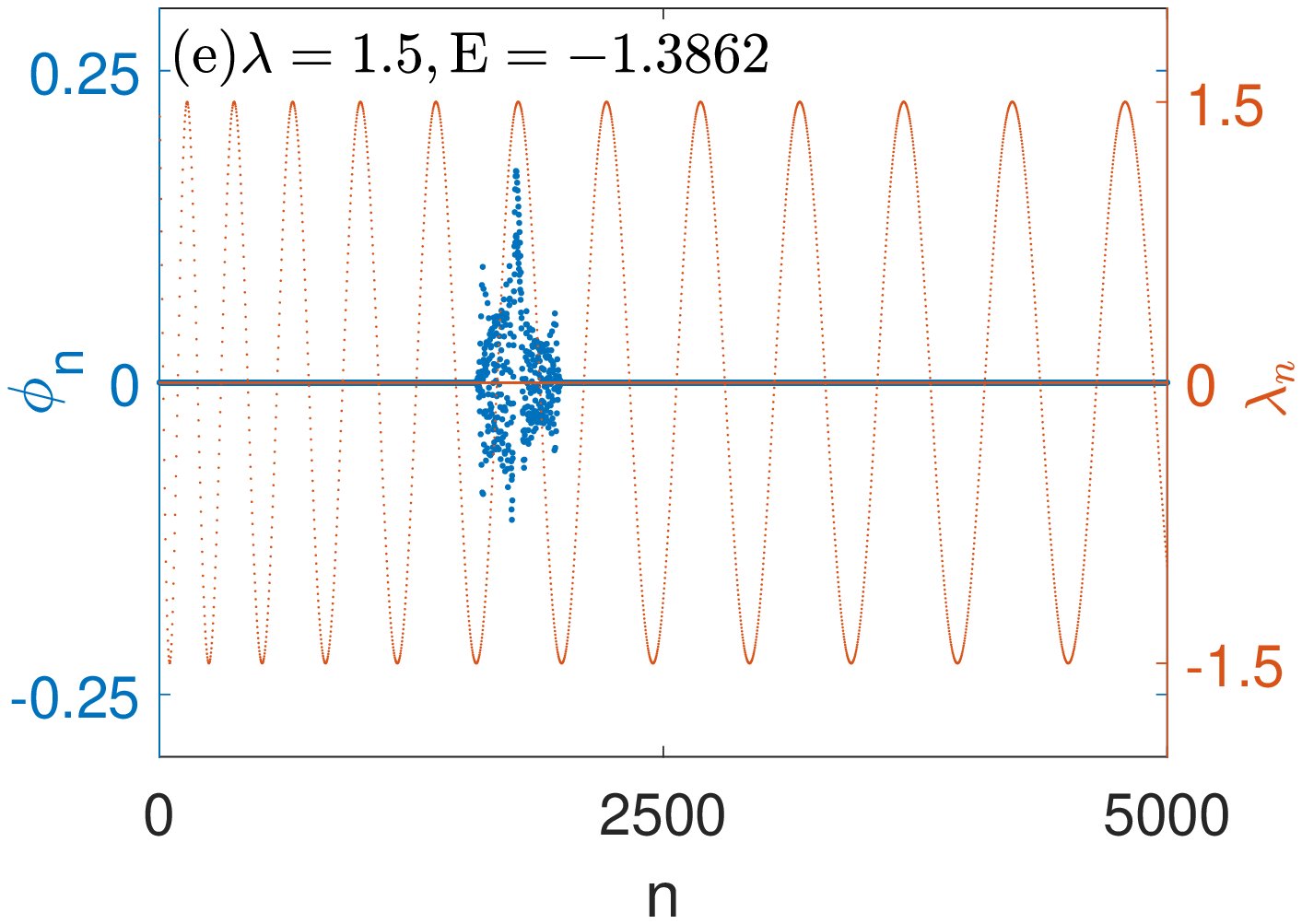}~
\includegraphics[width=1.69in]{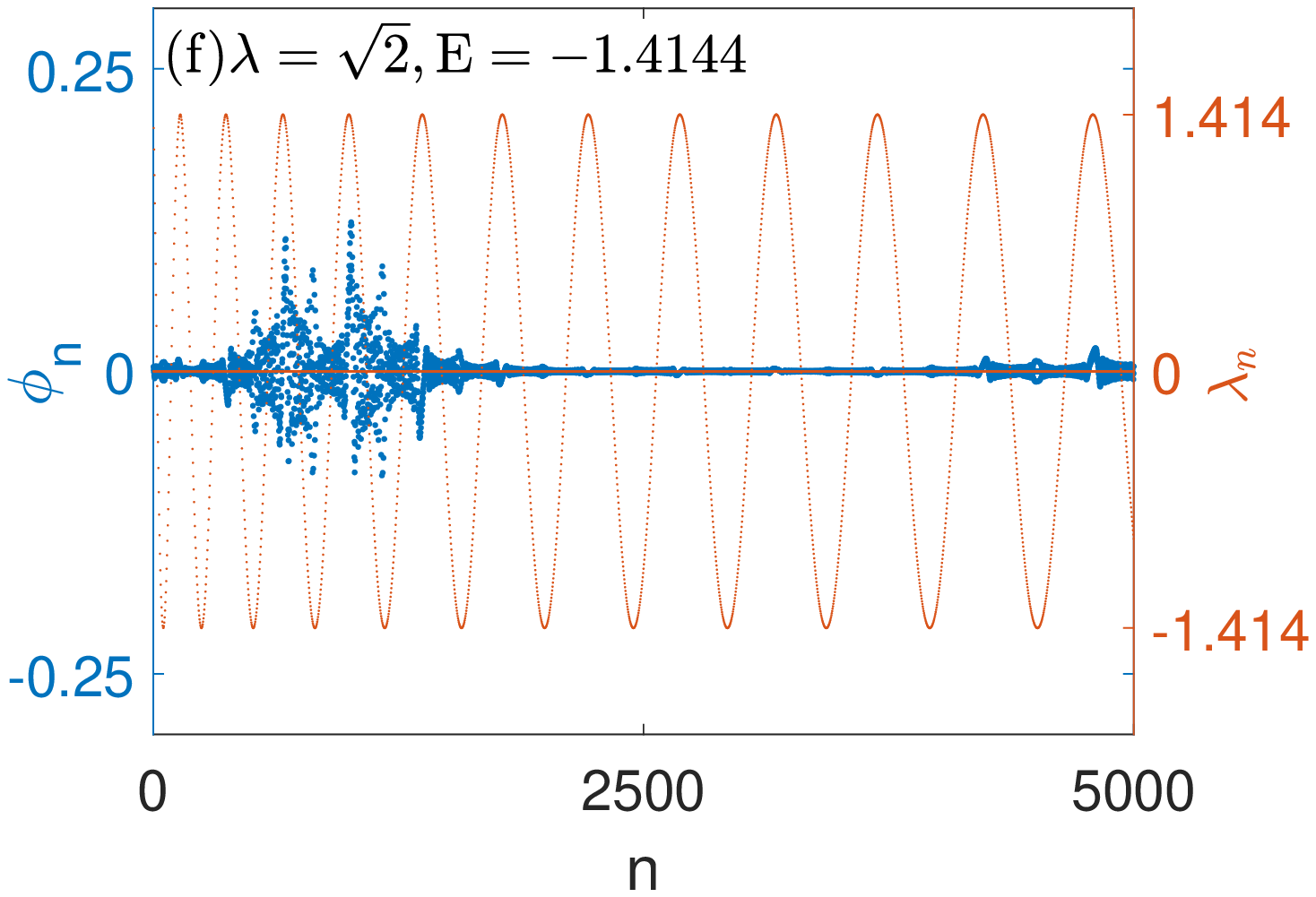}
\vspace{0.3cm}

\includegraphics[width=1.69in]{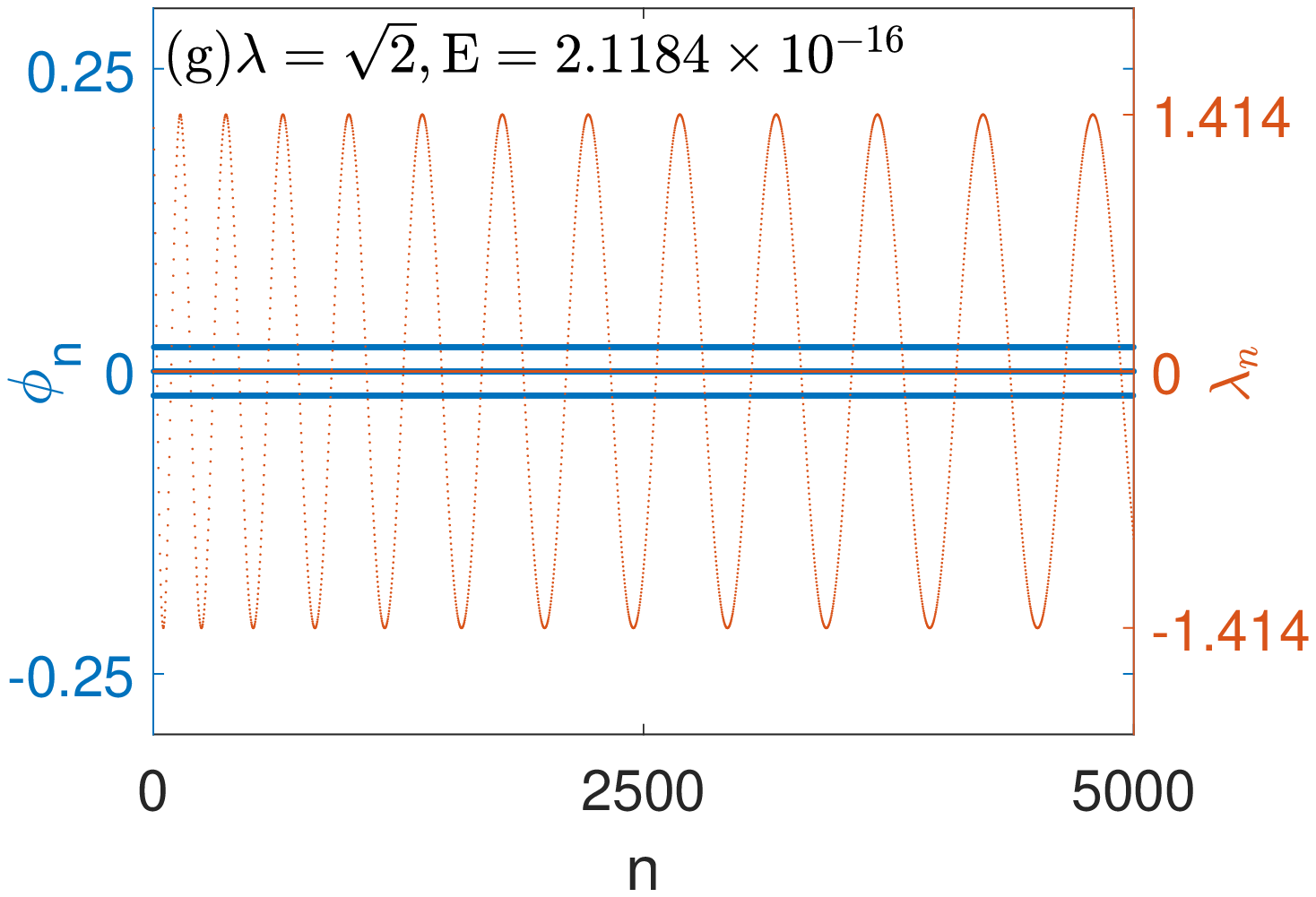}~~~~
\includegraphics[width=1.5in]{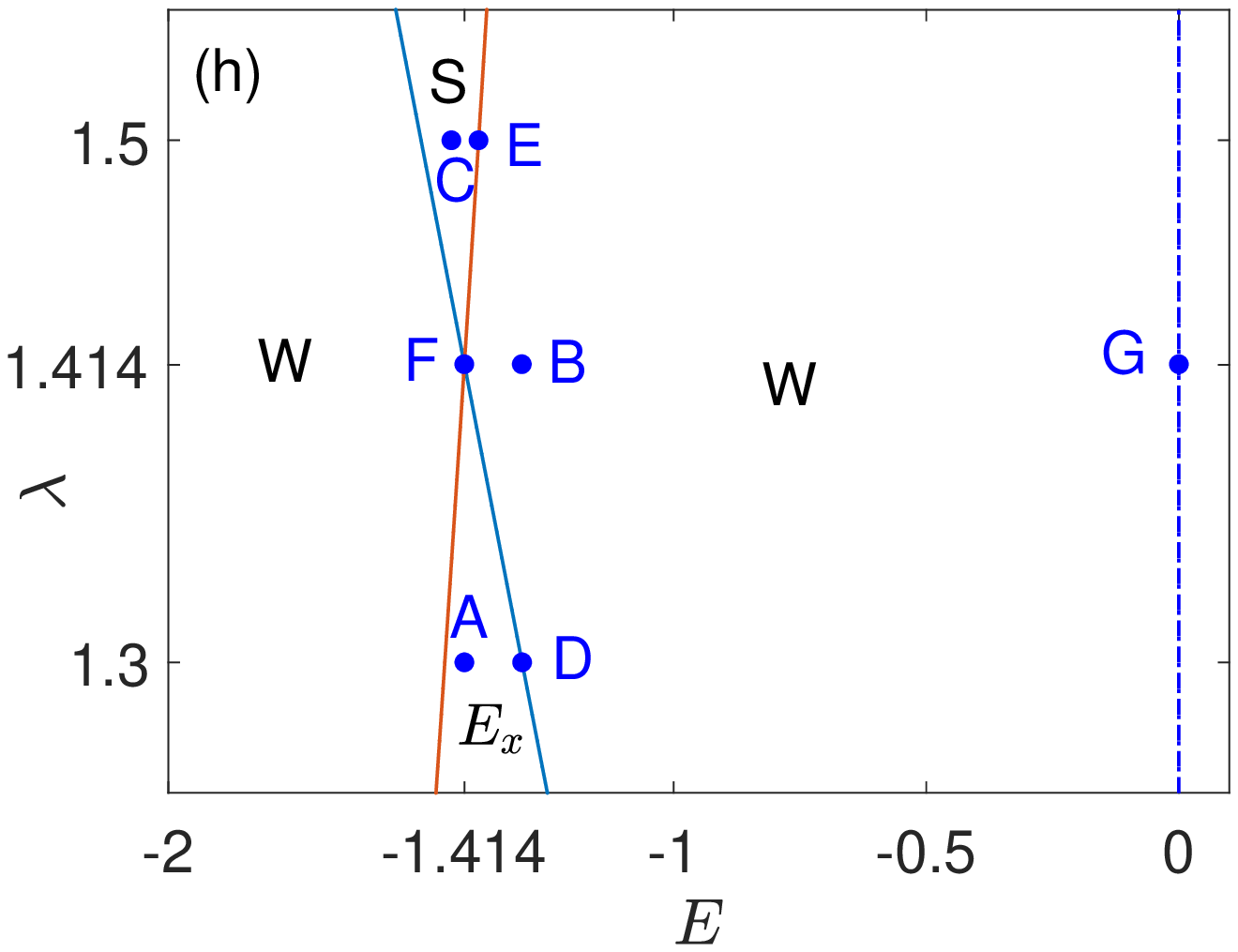}
\caption{For $[\kappa,m]=[2,0]$, some typical (a) extended, (b)weakly localized,(c) strongly localized and (d-g) critical eigenstates, which correspond to $A-G$ points in (f) the phase diagram. The potential $\lambda_n$ is also plotted in the red ground of (a-g).}\label{Fig5}
\end{figure}

Further, the phase diagram can also be directly reflected by the spatial distribution of wave functions. For $[\kappa,m]=[2,0]$, some typical eigenstates are plotted in Fig.\ref{Fig5} (a-g), and their positions in phase diagram labelled by $A-G$ points are shown in Fig.\ref{Fig5}(h). Fig.\ref{Fig5} shows the extended state spreads over the whole system, the strongly localized state is more localized than that of weakly localized one, and critical states [Figs.\ref{Fig5} (d) and (f)] at MEs are intermediately distributed in space. The localization property of the ``critical'' state [Fig.\ref{Fig5} (e)] at PME is intermediate between that of strongly localized and weakly localized states. The critical state [Fig.\ref{Fig5} (g)] at the band center is almost extended, which corresponds to that both $|\chi_{\lambda}|=1$ and $|\chi_{-\lambda}|=1$.

\emph{Numerical verification.}---
\begin{figure}[!htbp]
\includegraphics[width=1.65in]{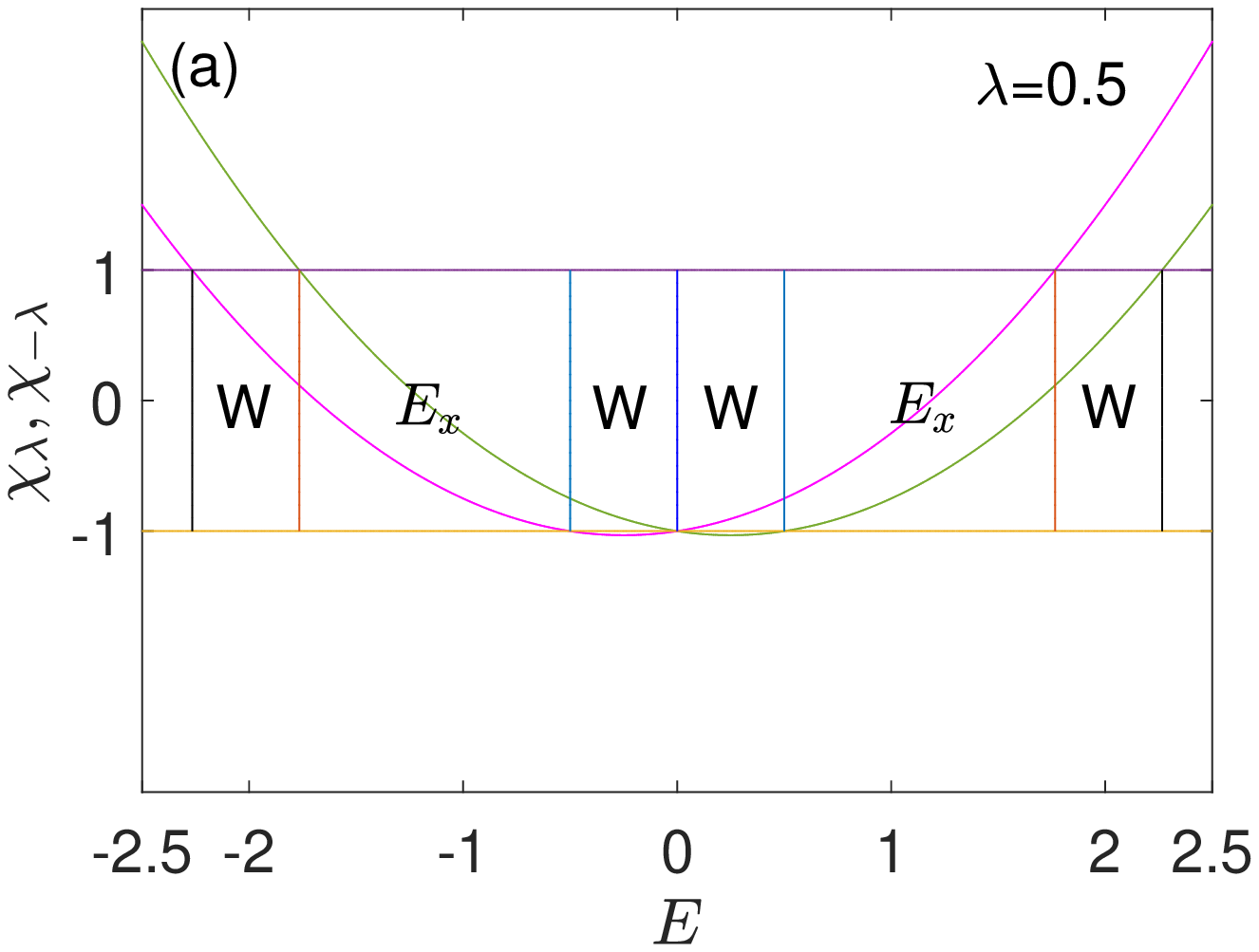}
\includegraphics[width=1.65in]{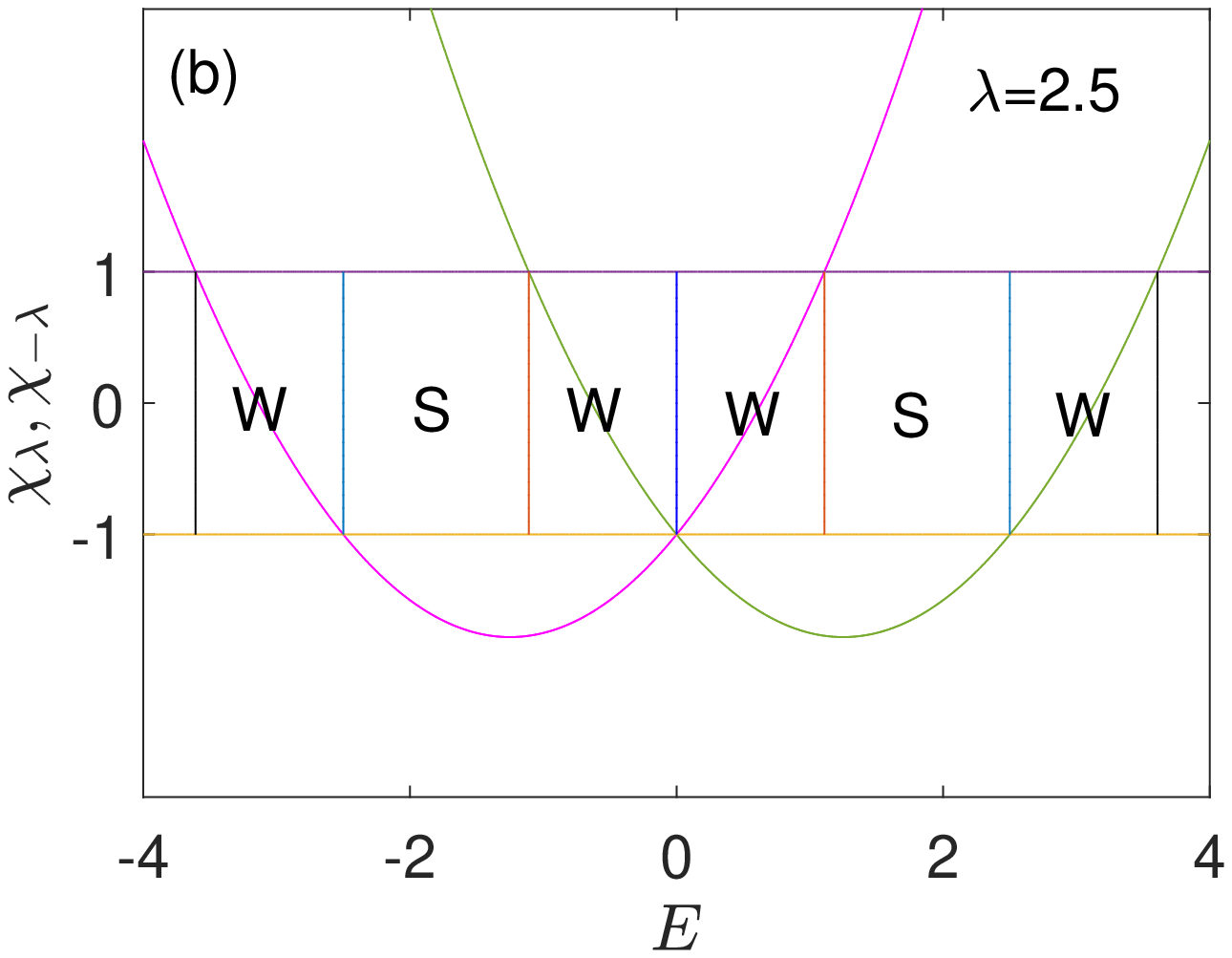}
\caption{For $[\kappa,m]=[2,0]$, the trace $\chi_{\lambda}$ (green line) and $\chi_{-\lambda}$ (magenta line) versus energies $E$ at (a) $\lambda=0.5$ and (b) $\lambda=2.5$, respectively. The vertical lines represent MEs and PMEs, and the horizon lines are for the functions $\chi=\pm1$.}\label{Fig6}
\end{figure}
Now we provide numerical verification about MEs and PMEs. Model parameters $[\kappa,m]=[2,0]$, $\lambda=0.5$ and  $\lambda=2.5$ are chosen as examples. Figs.\ref{Fig6}(a) and (b) show the variations of the trace $\chi_{\lambda}$ and $\chi_{-\lambda}$ with eigenenergies $E$. Based on the rule in Table I, extended ($E_x$), weakly localized ($W$) and stronger localized ($S$) regions are labelled. MEs and PMEs are represented by vertical lines.

\begin{figure}[!htbp]
\includegraphics[width=1.65in]{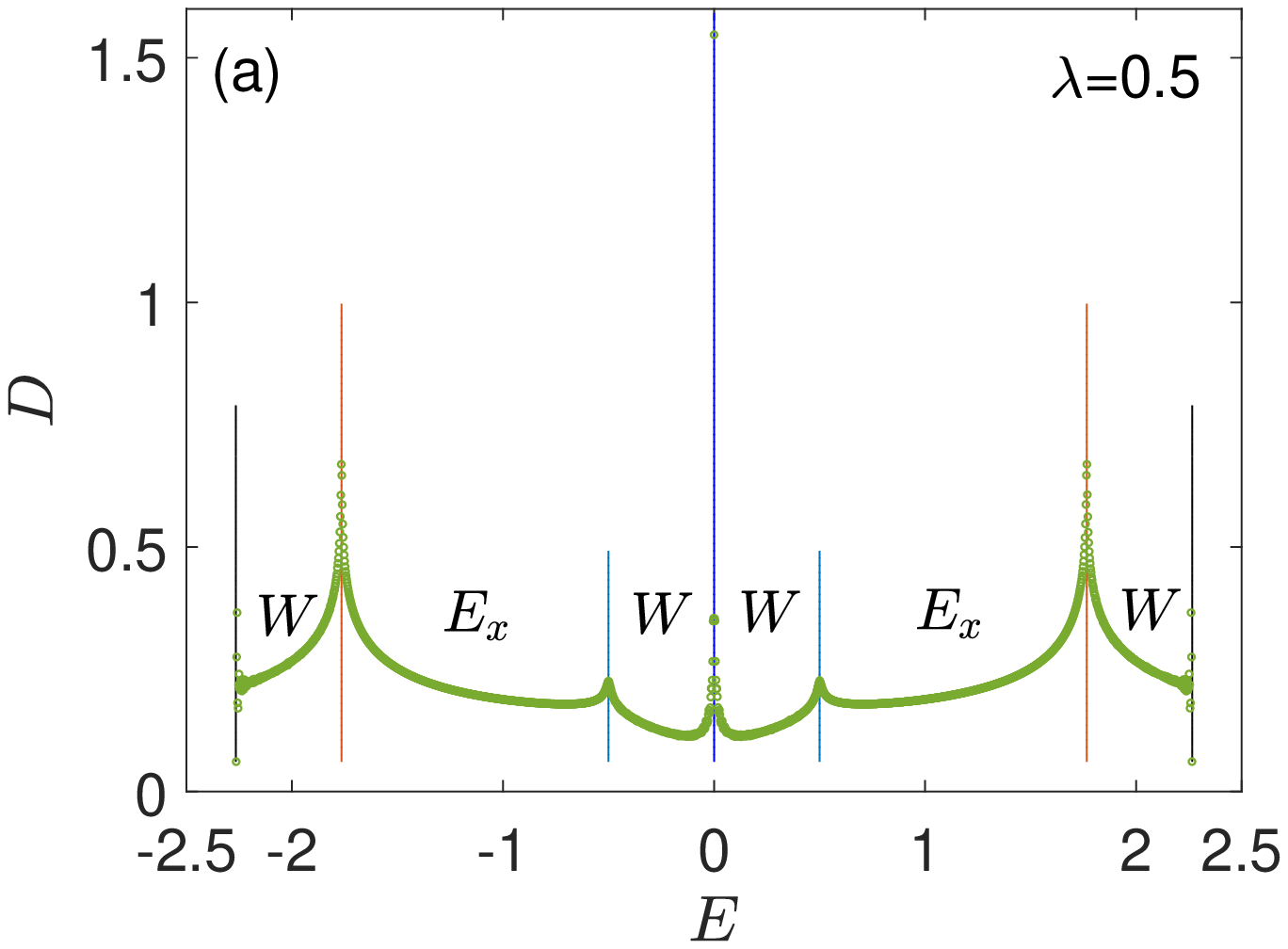}
\includegraphics[width=1.65in]{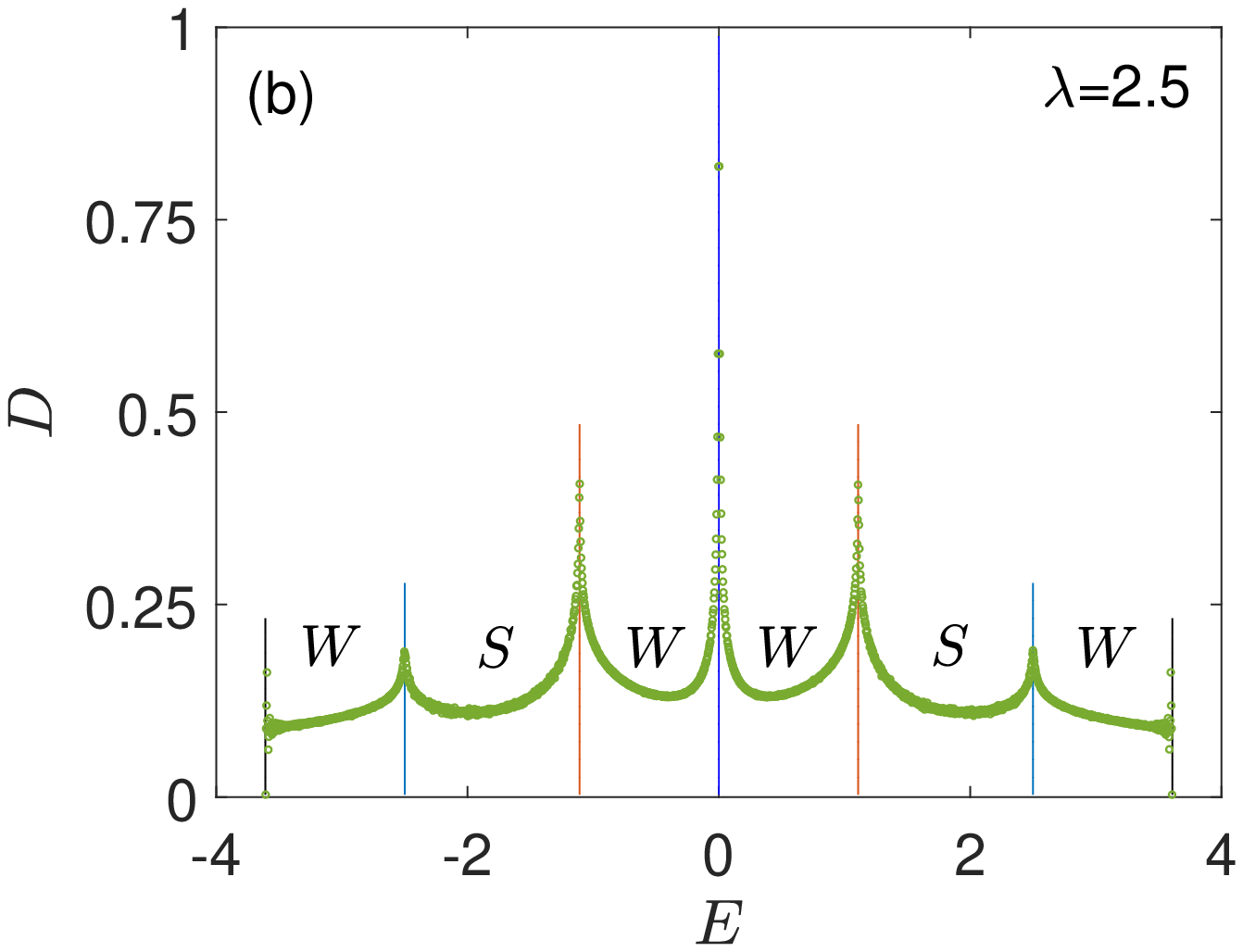}
\caption{For $[\kappa,m]=[2,0]$, the density of states $D$ (green circle) versus eigenenergies $E$ at (a) $\lambda=0.5$ and (b) $\lambda=2.5$, respectively. The vertical lines represent MEs and PMEs, and the system size $N=300000$.}\label{Fig7}
\end{figure}
To strengthen our findings, we calculate the density of states $D(E)$, which is defined by $D(E)=\sum_{n=1}^{N}\delta(E-E_n)$. Here $E_n$ is the $n$th eigenenergy. We use a method developed by Persson to compute all eigenvalues for large system with open boundary condition~\cite{PE}. It has been found that the singularity of the DOS can reflect MEs for quasiperiodic systems\cite{DA88,DA90}. Figs.\ref{Fig7} (a) and (b) show that there are sharp peaks in $D$ at MEs and PMEs, so they support our theoretical results.

\begin{figure}[!htbp]
\includegraphics[width=1.65in]{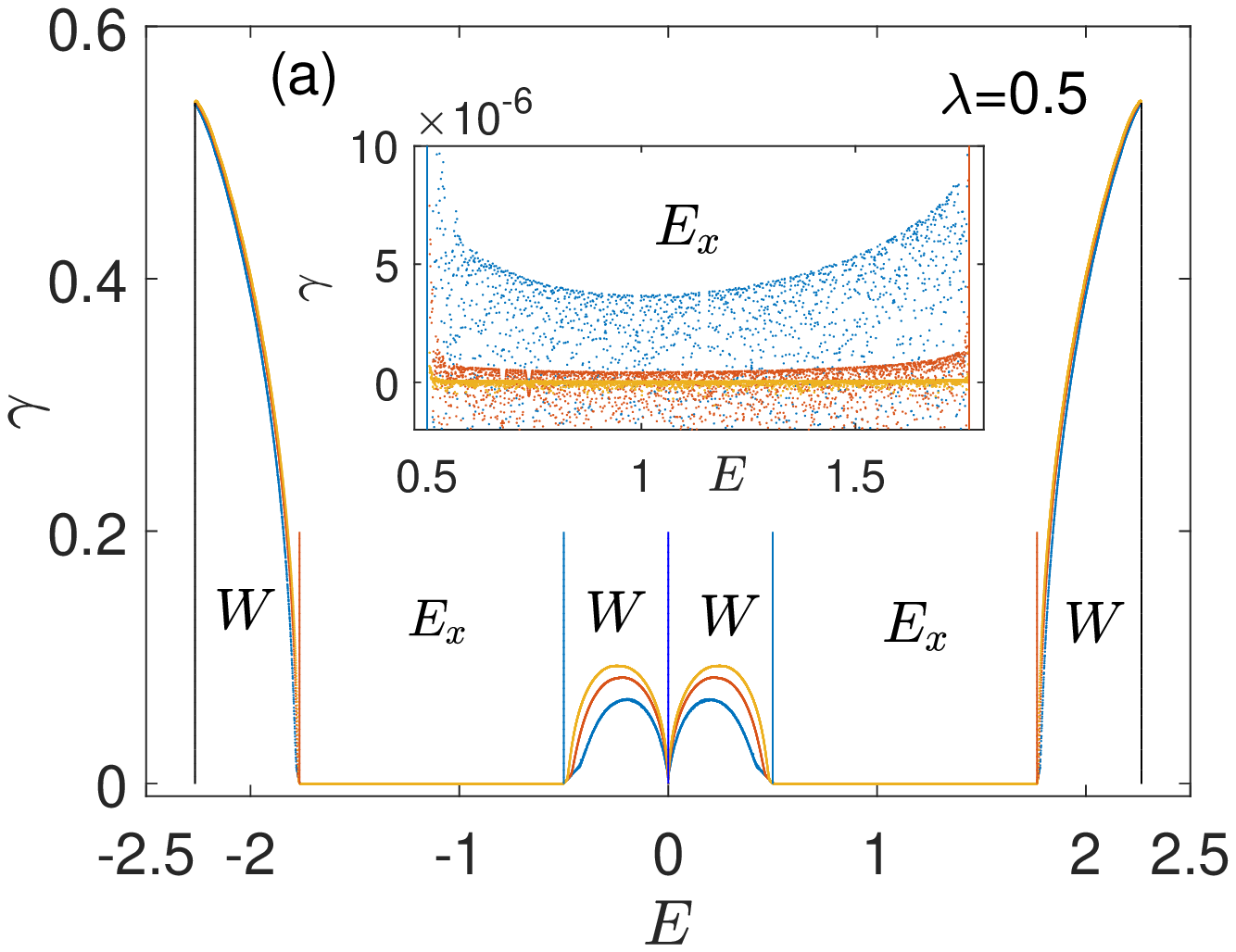}
\includegraphics[width=1.65in]{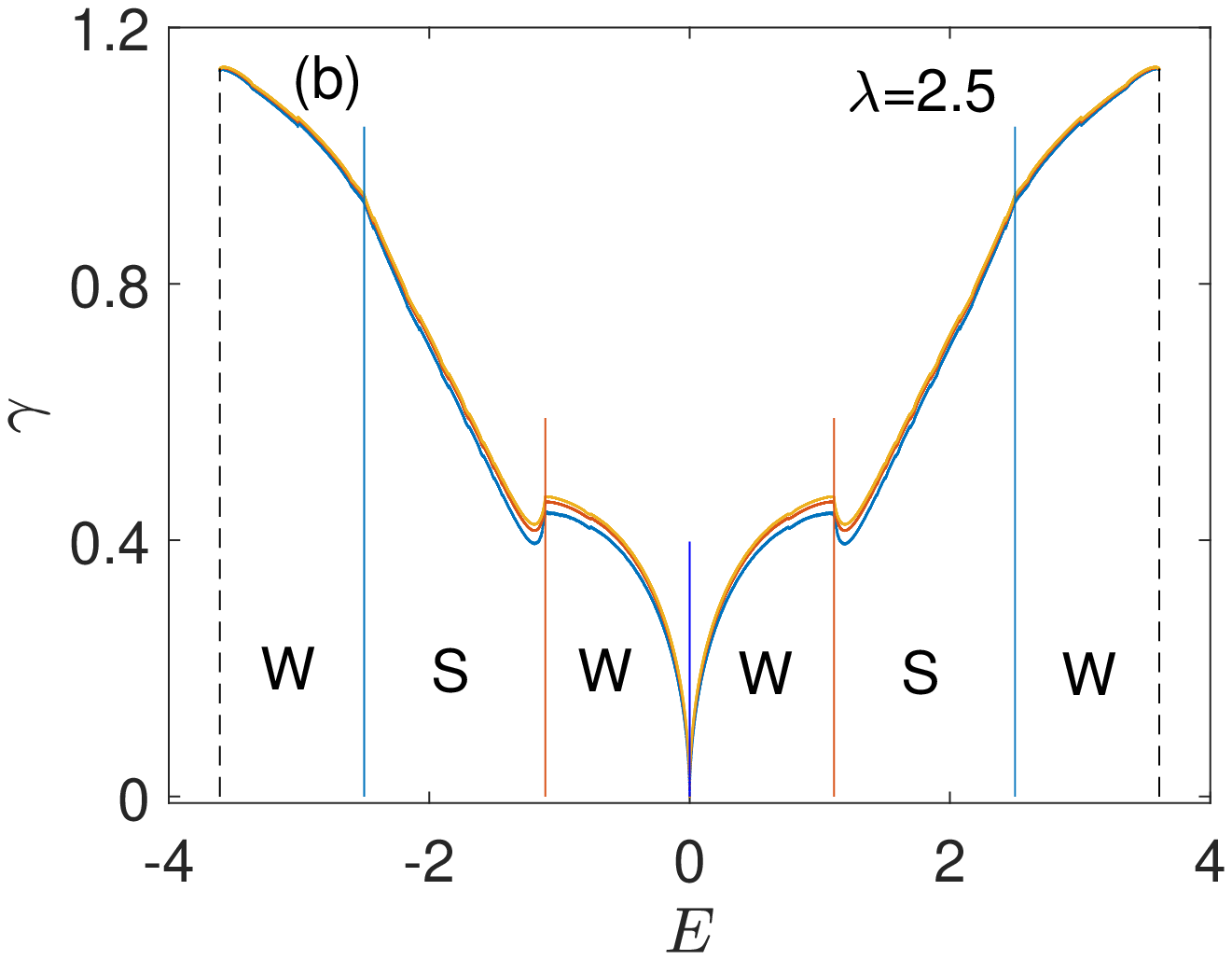}
\caption{For $[\kappa,m]=[2,0]$, the Lyapunov exponent $\gamma$ versus energies $E$ at (a) $\lambda=0.5$ and (b) $\lambda=2.5$, respectively. Inset in (a): Partial enlarger for one of extended regions. The vertical lines represent MEs and PMEs, and system sizes $N=10^5$ (blue dot), $10^6$ (red dot) and $10^7$ (yellow dot).}\label{Fig8}
\end{figure}

The energy-depended LE $\gamma(E)$ is often used to characterize the electronic localization properties. We use a numerically accurate renormalization scheme to
obtain it~\cite{FA92}. The LE is defined by
\begin{equation}
\gamma(E)=-\lim_{N\to\infty}[\frac{1}{N}\ln|t_{1N}^{eff}(E))], \label{EQ17}
\end{equation}
where $t_{1N}^{eff}(E)$ is the effective hopping integral between sites $1$ and $N$ when all the internal sites between them are properly decimated. In Figs.\ref{Fig8} (a) and (b) we plot the calculated LE as a function of the electron energies $E$ for $\pi\alpha=0.2$, $\nu=0.7$ and $\lambda=0.5$ and $2.5$, respectively. It is known that $\gamma=0$ for an extended state whereas $\gamma>0$ for a localized state. Fig.\ref{Fig8} shows that for extended states, $\gamma$ approaches zero; as shown in the inset of Fig.\ref{Fig8}(a),  the larger the system size is, the smaller the value of $\gamma$ is. For localized states, $\gamma>0$; the larger the system size is, the larger the value of $\gamma$ is. So the LE can well distinct extended regions from localized regions. Fig.\ref{Fig8} (b) shows there exist inflection points at PMEs, which can reflect the transitions between weakly localized phases and strongly localized ones. On the whole, numerical results about $\gamma$ well agree with our theoretical predictions.

\begin{figure}[!htbp]
\includegraphics[width=1.65in]{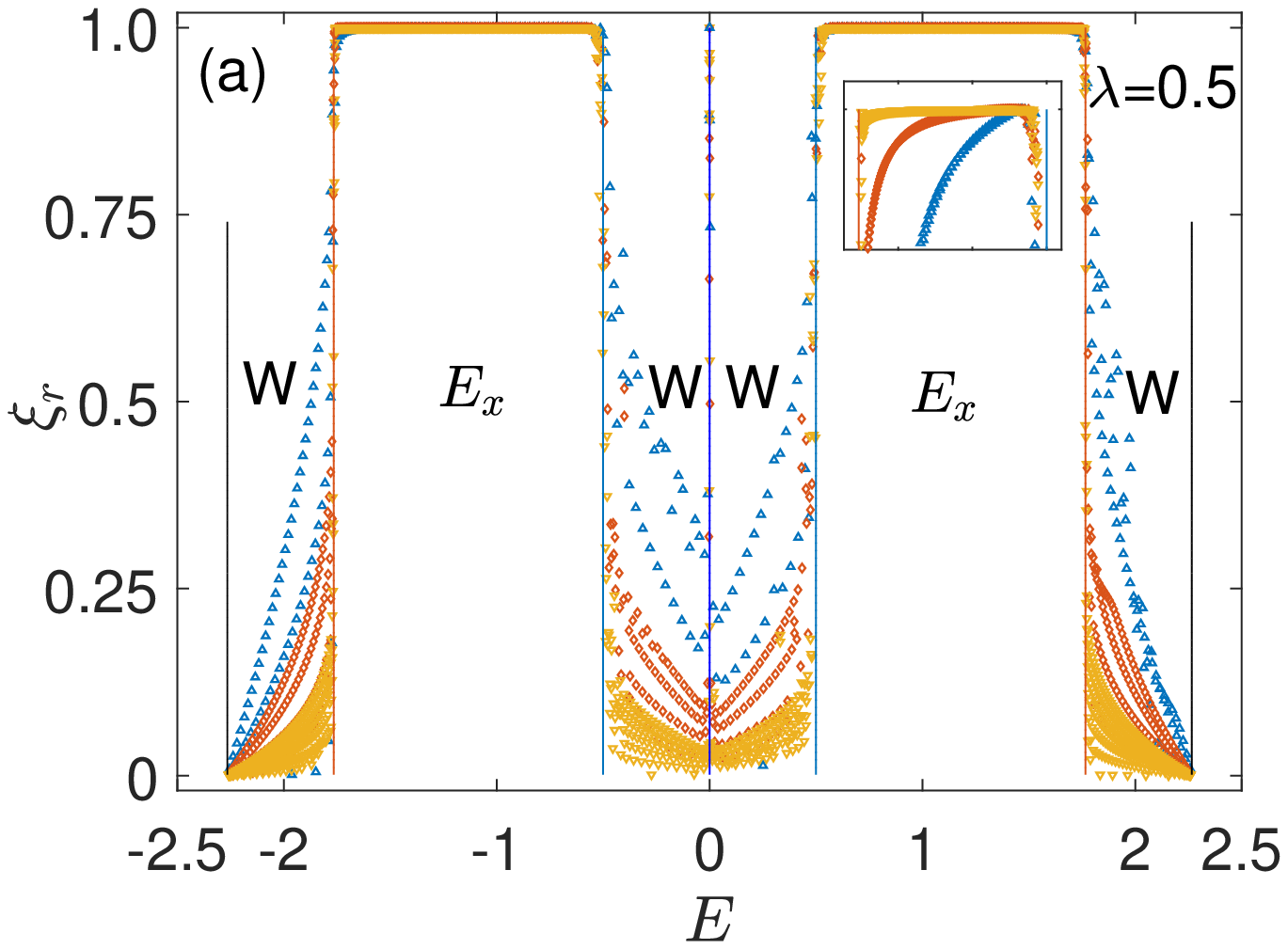}
\includegraphics[width=1.65in]{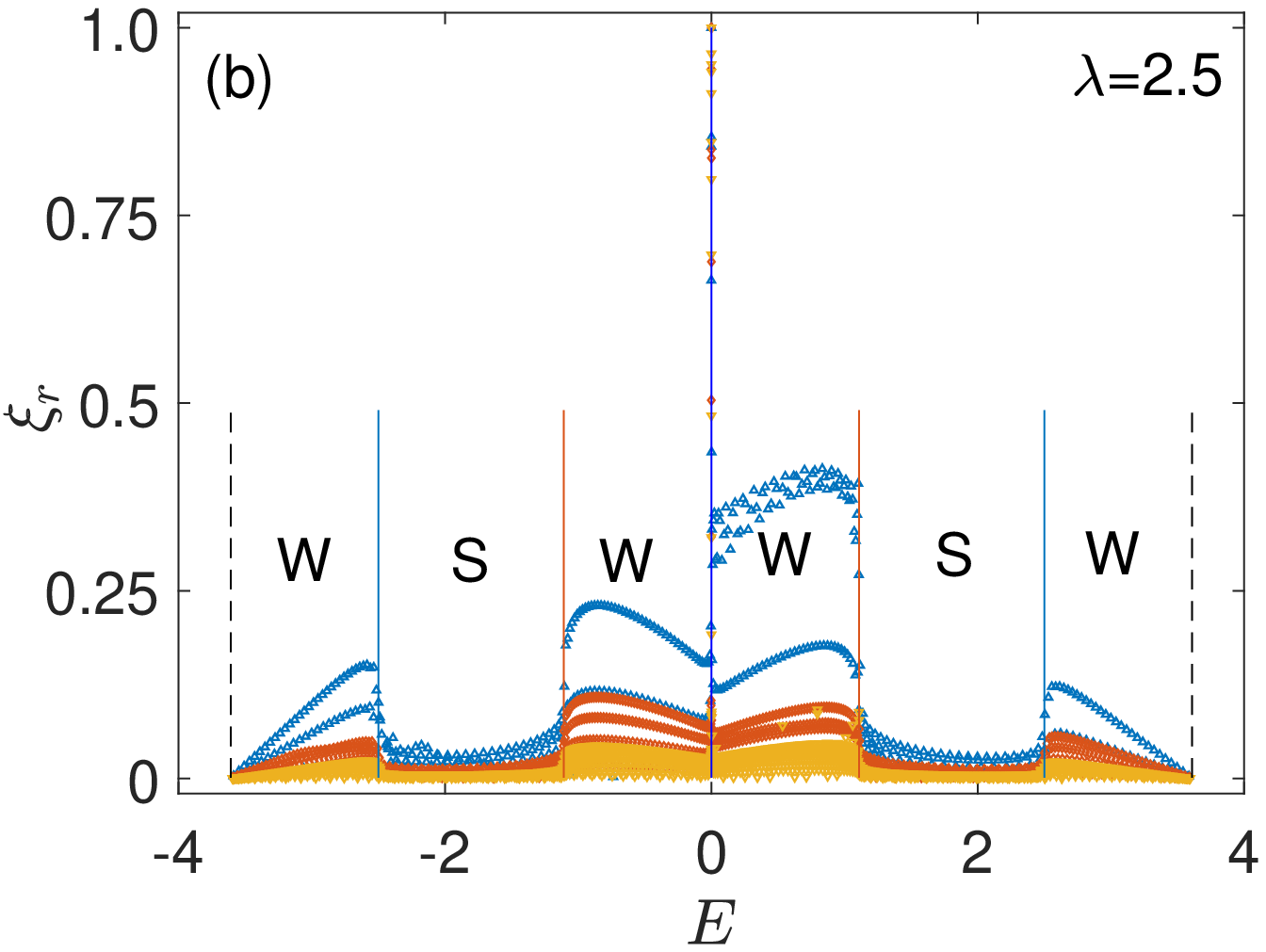}
\caption{For $[\kappa,m]=[2,0]$, the reduced localization tensor $\xi_r$ versus eigenenergies $E$ at (a) $\lambda=0.5$ and (b) $\lambda=2.5$, respectively. Inset in (a): Partial enlarger for one of extended regions. The vertical lines represent MEs and PMEs, and system sizes $N=500$ (blue upward triangle), $1000$ (red diamond) and $2000$ (yellow downward triangle).}\label{Fig9}
\end{figure}

However, as pointed out by Varga \emph{et al.}~\cite{VA92}, LE is not sufficient to characterize localized states beyond power-law shaped wave functions and  may lead to
erroneous. In Fig.\ref{Fig8}(b), LE $\gamma$ for states in strongly localized regions is smaller than that for states in weakly localized regions (near band edges), so it can not well reflect the localization properties of such localized states. Fortunately, for a periodic system the localization tensor (LT) $\xi$ is found to be an effective quantity that is able to distinguish metallic from insulating states~\cite{RE99,VA19}. For a single electron in 1D lattices, the LT is defined by
\begin{equation}
\xi=\langle\psi|q^{\dag}q|\psi\rangle-\langle\psi|q^{\dag}|\psi\rangle\langle\psi|q|\psi\rangle, \label{EQ18}
\end{equation}
where the complex position $q=\frac{N}{2{\pi}i}\sum_{n=1}^N[\exp(\frac{2{\pi}i}{N}n)-1]$, $N$ is the lattice size and $i=\sqrt{-1}$~\cite{VA19}. The maximal LT $\xi_{max}=\frac{N^2}{4{\pi}^2}$ for $|\psi\rangle=\frac{1}{\sqrt{N}}\exp(i\frac{2\pi j}{N}n)$, where $j$ is an integer
~\cite{VA19}. So we define a reduced LE
\begin{equation}
\xi_r=\xi/\xi_{max}. \label{EQ19}
\end{equation}
The calculations of $\xi_r$ are plotted in Fig.\ref{Fig9}. It shows that for extended states, $\xi_r$ approaches one; as shown in the inset of Fig.\ref{Fig9}(a), the larger the system size is, the larger the value of $\xi_r$ is. For localized states, $\xi_r$ is relatively small; the larger the system size is, the smaller the value of $\xi_r$ is. There are drastic changes in $\xi_r$ at MEs and PMEs. Therefore, it can well distinguish extended, localized and critical states. More importantly, Fig.\ref{Fig9}(b) shows on the whole, $\xi_r$ is relatively larger for weakly localized states than that for strongly localized ones. Therefore, the numerical results about $\xi_r$ are completely consistent with our theoretical predictions.

\emph{Conclusion.}---In this work, we propose a family of quasiperiodic mosaic lattice model base on a slowly varying potential. They have rich phase diagrams, including extended, weakly localized, strongly localized phases, mobility edges and pseudo-mobility edges. By using the asymptotic heuristic
argument and the theory of trace map of transfer matrix, we provide semi-analytical solutions. The semi-analytical results are in excellent agreement with the localization properties characterized by numerical calculations of the local density of states, the Lyapunov exponent, and the localization tensor. We expect our theoretical results to be directly observed in ultracold atoms and photonic waveguides.

\end{document}